\pdfoutput=1

\documentclass[11pt]{article}
\usepackage{jheppub}

\usepackage{amsmath}
\usepackage{amssymb}
\usepackage{graphicx,color,slashed,braket}
\usepackage{bbm}
\usepackage{ifpdf}
\usepackage{comment}
\usepackage{soul}

\usepackage{booktabs}
\usepackage{multirow}
\usepackage{makecell}
\usepackage{float}
\usepackage{verbatim}
\usepackage{caption}
\usepackage{cancel}
\usepackage{enumerate}
\usepackage{tcolorbox}
\usepackage[font=small,labelfont=bf]{caption}

\usepackage{color}
\definecolor{dark-gray}{gray}{0.20}
\definecolor{gray}{gray}{0.30}
\definecolor{light-gray}{gray}{0.80}
\definecolor{dark-red}{rgb}{0.7,0,0}
\definecolor{dark-green}{rgb}{0.1,0.4,0}
\definecolor{dark-blue}{rgb}{0.3,0.3,0.7}
\definecolor{light-blue}{rgb}{0.8,0.8,1}

\usepackage{tikz}
\usetikzlibrary{calc}

\usepackage{hyperref}
\usepackage{setspace}
\hypersetup{
	colorlinks=true,
	linkcolor=dark-blue,
	citecolor=dark-red,
	urlcolor=dark-green,
	linktoc=page
}
\newcommand{\be}{\begin{equation}}
\newcommand{\ee}{\end{equation}}

\def\be{\begin{equation}}
\def\ee{\end{equation}}
\def\bea{\begin{eqnarray}}
\def\eea{\end{eqnarray}}

\newcommand{\der}{\partial}
\newcommand{\de}{\mathrm{d}}

\newcommand{\e}{\mathrm{e}}
\newcommand{\I}{\mathrm{i}}

\newcommand{\dd}{\mathrm{d}}

\newcommand{\Mpl}{M_\text{Pl}}

\renewcommand{\Re}{\text{Re}\,}

\allowdisplaybreaks

\usepackage[backgroundcolor=white!95!red, linecolor=cyan, bordercolor=cyan, textsize=footnotesize]{todonotes}

\title{AdS scale separation and the distance conjecture}

\author{Gary Shiu$^1$, Flavio Tonioni$^{1}$, Vincent Van Hemelryck$^{1,2}$, Thomas Van Riet$^2$}
\affiliation{$^1 $ Department of Physics, University of Wisconsin-Madison, 1150 University Avenue, \\ Madison, WI 53706, USA}
\affiliation{$^2 $ Instituut voor Theoretische Fysica, K.U. Leuven,
Celestijnenlaan 200D, B-3001 Leuven, Belgium}

\emailAdd{shiu@physics.wisc.edu}
\emailAdd{tonioni@wisc.edu}
\emailAdd{vincent.vanhemelryck@kuleuven.be}
\emailAdd{thomas.vanriet@kuleuven.be}

\notoc

\abstract{It has been argued that orientifold vacua with fluxes in type IIA string theory can achieve moduli stabilisation and arbitrary decoupling between the AdS and KK scales upon sending certain unconstrained RR-flux quanta to infinity. In this paper, we find a novel scalar field in the open-string sector that allows us to interpolate between such IIA vacua that differ in flux quanta and find that the limit of large fluxes is nicely consistent with the distance conjecture. This shows that the massive IIA vacua pass an important Swampland criterion and suggests that scale-separated AdS vacua might not be in the Swampland. Our analysis also naturally suggests a flux analogue of ``Reid's fantasy'' where flux vacua that differ in quantised flux numbers can be connected through trajectories in open-string field space and not just via singular domain walls.}

\begin{document}

\maketitle

\newpage
\tableofcontents
\newpage
\section{Introduction}
One of the most sought-after features of anti-de Sitter (AdS) vacua in string theory is the notion of scale separation, defined as the property that
\be
\frac{L_\text{KK}}{L_\text{AdS}} \ll 1\,.
\ee
Here, by $L_\text{KK}$ we denote the Kaluza-Klein length scale, which is defined as the inverse mass scale for Kaluza-Klein excitations, and by $L_\text{AdS}$ we mean the AdS length defined in the usual way through the cosmological constant $\Lambda = - 3 / L_\text{AdS}^2$. Scale separation might be a necessary condition for constructing semi-realistic vacua as the compact dimensions are then invisible to low-energy observers. Achieving scale separation in string-theory constructions is particularly hard and only a few proposals benefit from this feature \cite{Derendinger:2004jn, DeWolfe:2005uu, Camara:2005dc, Cribiori:2021djm, Farakos:2020phe, VanHemelryck:2022ynr, Kachru:2003aw,Balasubramanian:2005zx}, although several works claim it is even impossible in some cases \cite{Lust:2019zwm, Collins:2022nux, Lust:2022lfc, Cribiori:2022trc, Montero:2022ghl}.
Alternatively, one can think of interesting phenomenological scenarios where the absence of scale separation is a positive feature instead \cite{Montero:2022prj}. 

In the context of holography, one usually does not not encounter scale-separated AdS vacua, but the question of what the CFT duals to scale-separated vacua would be has been investigated to some extent; see e.g. refs. \cite{Polchinski:2009ch, Alday:2019qrf, Collins:2022nux} for general remarks and refs. \cite{Aharony:2008wz,Conlon:2018vov, Conlon:2020wmc, Conlon:2021cjk, Apers:2022tfm, Apers:2022vfp, Apers:2022zjx, Quirant:2022fpn, Plauschinn:2022ztd, Lust:2022lfc} for studies specific to models in the string phenomenology literature. 

At the time of writing, the proposed constructions of scale-separated vacua falls roughly into two broad classes of suggested mechanisms.
\begin{itemize}
    \item On the one hand, there are quantum-corrected no-scale vacua, such as the KKLT scenario \cite{Kachru:2003aw}, the Large Volume Scenario (LVS) \cite{Balasubramanian:2005zx} and the like. Whether these AdS vacua are under control is a long-standing debate that has intensified in recent years. In particular, there seems to be a tension in finding controlled vacua that can be uplifted to de Sitter using anti-branes; see ref. \cite{Danielsson:2018ztv} for an overview and refs. \cite{Gao:2020xqh, Bena:2020xrh, Emelin:2020buq, Junghans:2022exo, DallAgata:2022abm, Farakos:2020wfc, Junghans:2022kxg, Blumenhagen:2022dbo, Lust:2022lfc} for recent papers pointing to specific problems with the existing constructions and see refs. \cite{Demirtas:2021nlu, Demirtas:2021ote} for arguments why sufficient control can be achieved (for AdS vacua before the uplift). Recently it has been appreciated that quantum corrections in the form of Casimir energies can be relevant for achieving scale separation \cite{DeLuca:2022wfq}.
    
    \item On the other hand, one has classical flux vacua with orientifold planes, such as DGKT-like vacua \cite{Derendinger:2004jn, DeWolfe:2005uu, Camara:2005dc, Cribiori:2021djm, Farakos:2020phe, VanHemelryck:2022ynr}.\footnote{See also refs. \cite{Acharya:2002kv, Petrini:2013ika, Caviezel:2009tu, Caviezel:2008ik, Richard:2014qsa, Tsimpis:2012tu, Buratti:2020kda, Lust:2020npd, Font:2019uva, Basile:2022ypo, Marchesano:2020uqz,  Emelin:2021gzx, Andriot:2022yyj} for various investigations as to whether scale separation is possible in classical flux models.} Note that the IIB vacua of this class \cite{Petrini:2013ika, Caviezel:2009tu} do not seem under computational control because of certain cycles that remain of stringy size \cite{Cribiori:2021djm}. 
    The orientifolds in the classical flux vacua were argued to be the crucial ingredients \cite{Gautason:2015tig, DeLuca:2021mcj, DeLuca:2021ojx} for achieving scale separation, which requires compact manifolds with positive-definite Ricci tensor whose KK scale can become large at fixed curvature radius. The latter has been conjectured to be impossible \cite{Collins:2022nux}, although the lift of the scale-separated IIA vacua without Romans mass is supposed to lift to Freund-Rubin type vacua without negative tension sources which are nonetheless scale-separated \cite{Cribiori:2021djm}. 
\end{itemize}

None of the above quoted examples are considered as fully established string-theoretic vacua. For the first class, this is related to the absence of a parametric regime: not a single quantity can be sent to extreme values, mostly due to tadpole conditions, which are more restrictive than originally envisaged \cite{Bena:2020xrh}. This is at the heart of the debate and recent holographic arguments suggest that perhaps these vacua are not controlled \cite{Lust:2022lfc}. Concerning the classical orientifold vacua (second class), one criticism has been that when a Romans mass is involved one cannot control the orientifold vacua \cite{McOrist:2012yc}. This criticism has been shown not to hold in explicit examples \cite{Baines:2020dmu} and, furthermore, there exist vacua in massless IIA as well \cite{Cribiori:2021djm}. A second, related, source of criticism has been the incomplete treatment of the orientifold backreaction \cite{Banks:2006hg}. The effective description of ref. \cite{Grimm:2004ua} has been constructed in the coarse-grained picture that smears out the orientifold singularities \cite{Acharya:2006ne, Blaback:2010sj}. This can also be understood in explicit examples showing how this is harmless when the zero mode of the string coupling can be tuned small and the volume is large enough \cite{Baines:2020dmu}. But, more importantly, partial insight on the backreaction of the localised orientifolds exist as a perturbative series for which the leading correction to the smeared picture has been computed \cite{Marchesano:2020qvg, Junghans:2020acz} (see also refs. \cite{Cribiori:2021djm, Marchesano:2022rpr, Emelin:2022cac}). 

In the absence of a full backreaction picture, some doubts on the consistency of these vacua perhaps remain and these have led to a Swampland conjecture known as the AdS distance conjecture (ADC) \cite{Lust:2019zwm} (see also refs. \cite{Cribiori:2021gbf, Castellano:2021yye, DallAgata:2021nnr}). For supersymmetric (SUSY) AdS vacua, its strong version states that $L_\text{AdS}\sim L_\text{KK}$ and so scale separation is not achievable. If correct, the type IIA vacua are in the Swampland.  Interestingly the ADC can be tied to the magnetic Weak Gravity Conjecture (WGC) for AdS vacua with extended supersymmetry \cite{Cribiori:2022trc}. Since the WGC is on a firm footing, this gives confidence for the absence of scale separation when there is extended supersymmetry. This is consistent with the fact that all known proposals for scale separation have minimal or no supersymmetry. Another, yet related, observation that connects the absence of scale separation with firmly established Swampland principles appeared in ref. \cite{Montero:2022ghl}, where it was shown that extreme scale separation (no KK sector) for sufficiently supersymmetric AdS vacua implies that the R-symmetry becomes global in the large-$N$ limit. 

The argument behind the ADC relies on an extension of the swampland distance conjecture (SDC) of ref. \cite{Ooguri:2006in}, for which quite some confidence has been built up over the last years. The distance conjecture quantifies how an effective field theory breaks down when moving far away in scalar field space. In Planck units, at large geodesic distance $\Delta$ in field space from the original vacuum of the EFT, the mass scale $m$ of a tower of modes becomes lighter as an exponential of the negative of such a distance, i.e.
\be
    m \sim m_0 \, \e^{- \beta \Delta} \,,
\ee
with $\beta$ an order-one number \cite{Etheredge:2022opl}. Then, ref. \cite{Lust:2019zwm} took the point of view that perhaps this conjecture also holds for distances travelled in general field space, not just scalar field space. When applied to the metric field, one then finds the ADC. However, more evidence is needed to substantiate such an extension of the distance conjecture. This motivates us to find a scalar field that can interpolate between IIA vacua with different flux numbers. Naively, this cannot be done since changing flux numbers occurs through Brown-Teitelboim domain-wall transitions. Yet, for IIB vacua, it was known \cite{Kachru:2002gs} that this can also be described by a motion of an open-string field and it was subsequently used to study the distance conjecture \cite{Scalisi:2020jal}. In this paper we will also find such a field for the IIA vacua and show that the motion in field space towards scale separation is exactly consistent with the standard distance conjecture.

The rest of this paper is organised as follows. To start, in section \ref{sec: DGKT review} we highlight the key aspects of massive IIA 4d vacua. Then, in section \ref{sec: DGKT interpolating D4-brane} we provide arguments to explain the role of D4-branes in interpolating among different such vacua and in section \ref{sec: open-string SDC} we show that the scalar controlling the motion of these D4-branes realises the swampland distance conjecture. In section \ref{sec: other classes of vacua} we discuss analogous mechanisms in other classes of flux vacua. To conclude, in section \ref{sec: conclusions} we summarise our results.

\section{Massive IIA flux vacua with scale separation} \label{sec: DGKT review}
Let us present a lightning review of the 10d description of the massive IIA 4d vacua \cite{DeWolfe:2005uu}. Consider a Calabi-Yau three-fold and its canonical K\"ahler 2-form $J$ and holomorphic 3-form $\Omega$. In the 10d string frame and in string units, given a real number $m$, the Ansatz \cite{Lust:2004ig, Grana:2006kf, Behrndt:2004mj}
\begin{subequations}
\begin{align}
    & F_0 = 5 g_s^{-1} m \,, \label{eq:F0_DGKT} \\
    & F_4 = \frac{3}{2} m g_s^{-1} J\wedge J\,,\\
    & H = 2m \, \Re\Omega \,, \label{eq:H_DGKT}
\end{align}
\end{subequations}
solves the 10d equations of motion if the Bianchi identities with smeared O6-sources are satisfied \cite{Grana:2004bg, Acharya:2006ne},\footnote{The O6-planes are unavoidable as one requires a net negative tension \cite{Gautason:2015tla}. This means that the Calabi-Yau has to allow for the right kind of involutions.} and provides a 4d AdS geometry with a string-frame cosmological constant given (again in string units) by
\be
\Lambda = -\frac{3}{L_\text{AdS}^2} = -3m^2\,.
\ee
For the sake of simplicity we will consider toroidal orbifolds when an explicit geometry is needed and further simplify to so-called isotropic solutions. In terms of Cartesian coordinates on a torus, this means that one can write
\begin{subequations}
\begin{align}
    & J = L^2 (\dd x^1\wedge \dd x^2 + \dd x^3 \wedge \dd x^4 + \dd x^5\wedge \dd x^6) \,, \\
    & \Omega = L^3 (\I \, \dd x^1 + \dd x^2) \wedge (\I \, \dd x^3 +\dd x^4) \wedge (\I \, \dd x^5 + \dd x^6) \,.
\end{align}
\end{subequations}
The Romans mass $F_0$ and the NSNS-flux $H$ are restricted by the RR-tadpole condition coming from integrating the Bianchi identity
\be
\label{eq:F2BI_DGKT}
\dd F_2 = F_0 H + \delta_\text{O6}\,,
\ee
where $\delta_\text{O6}$ represents the local charge contribution of the O6-planes. All the equations \eqref{eq:F0_DGKT}-\eqref{eq:H_DGKT} and \eqref{eq:F2BI_DGKT} are subject to a scaling symmetry and since Romans mass and $H$-flux quanta are order-one and fixed, we can identify the scalings
\begin{equation}
    g_s \sim N^{-3/4} \,, \qquad
    L \sim N^{1/4} \,, \qquad
    L_\text{AdS} \sim N^{3/4} \,.
\end{equation}
Here $N$ is the $F_4$-flux quantum, which is unconstrained by tadpoles. Therefore, we can consistently consider the large-$N$ limit. Note in particular that the Planck mass diverges in this limit, as $\Mpl^2 \sim g_s^{-2}L^6 \sim N^3$, but that we achieve scale separation as\footnote{We take $L_\text{KK}=L$ as an estimate.}
\be
\frac{L_\text{KK}}{L_\text{AdS}} \sim N^{-1/2}\,.
\ee
The same limit seems to put the solution consistently in the regime where classical 10d supergravity is a valid description. Finally, note that the vacuum energy density $V = L^6 \Lambda = \Mpl^2 \Lambda_4$ remains parametrically smaller than the Planck scale but in fact diverges (in string units) as
\be
V \sim N^{-9/2} \Mpl^4 \sim N^{-3/2} \Mpl^2 \sim N^{3/2}.
\ee
Every different integer $N$ gives rise to a different vacuum solution. In the next section we discuss how one can interpolate between the different flux vacua.

\section{Interpolating between flux vacua} \label{sec: DGKT interpolating D4-brane}
The $F_4$-flux quantum $N$ is clearly the most important integer that describes these vacua. Its value cannot be changed in a dynamical process since these vacua are supersymmetric and hence stable. In other words, the D4-brane domain walls that could lower or increase the flux have charge equal to tension and hence are marginally stable. This is depicted in fig. \ref{bubble}, where inside the wall the flux quantum $N_1$ is different from the outside, where it is $N_2$. The difference in flux is compensated by the wall charge and the difference in cosmological constant by its tension. For SUSY AdS vacua, this bubble does not move outward. For the non-SUSY IIA solutions it was not clear whether they could decay at all \cite{Narayan:2010em}, but a recent analysis \cite{Marchesano:2021ycx,Marchesano:2022rpr} at least has shown that the subset of vacua that invokes some D6-branes to cancel the RR-tadpole are unstable and decay to non-SUSY solutions without D6-branes, for which the decay is not yet known. This decay is not through the D4-branes, though.

Our interest is in the  SUSY IIA vacua where the D4-brane domain walls wrap holomorphic 2-cycles $\Sigma_2$ which are Poincaré-dual to the 4-form flux. These domain walls are formal transitions between the vacua, but as mentioned above, do not occur dynamically. Since the 4-form flux is proportional to $J \wedge J$, the D4-branes wrap 2-cycles with a volume form proportional $J$, which calibrates that 2-cycle.

\subsection{Unwinding the flux without domain walls for IIA vacua}

\begin{figure}[h]
    \centering
    \begin{tikzpicture}[scale=0.25]
    
    \draw[thick, draw=gray, fill=gray!10!white] (0,0) circle (5);
    
    \node[] at (0,0){$N_1$};
    \node[] at (-6,3.5){$N_2$}[below];
    \node[below right] at (4,-3){};
    
    \end{tikzpicture}
    \caption{There are two possible flux discharge processes and both can be represented with the same picture. In one case the bubble is codimension-1 inside 4d space, with the D4-brane wrapping the holomorphic cycle $\Sigma_2$ and in the other case it is codimension-1 inside $\Sigma_2$ with the D4-brane filling 4d spacetime.} \label{bubble}
\end{figure}
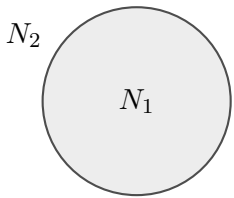

In what follows, we want to consider other objects that still interpolate between different flux integers but that are not domain walls inside the 4d spacetime AdS$_4$. Instead, we look at D4-branes that fill 4d space and wrap a contractible cycle inside the holomorphic 2-cycle $\Sigma_2$. To picture that, we can again rely on fig. \ref{bubble}, but now with the background representing $\Sigma_2$ instead of AdS$_4$. Since the object is codimension-1 inside $\Sigma_2$, it must be that the 4-form flux differs inside and outside. An observer in AdS$_4$ is not localised inside the extra dimension but will see a different vacuum once the bubble has been moved completely over the 2-cycle.

Since this might seem somewhat unconventional, we explain it in some more detail. Imagine that we change the topology of our AdS$_4$ solution by working in the Poincaré patch and make the spatial slice into a 3-torus. Then we still have the same solution to the equations of motion -- only the topology differs -- but the field equations are blind to this. From the point of view of the D4-brane, all compact directions are equivalent: as long as the D4-branes are transverse to the 4-cycle filled with $F_4$-flux, they are codimension-1 in the six directions transverse to that 4-cycle, and one side of the D4-brane will have different $F_4$-flux from the other side. This can be understood from investigating the $F_4$-form Bianchi identity, which for such a setup takes the form
\be
    \dd F_4 = Q_\text{D4} \, \delta(y,z^1,z^2,z^3,z^4) \, \dd y \wedge  \epsilon_4 \,,
\ee
with $\epsilon_4$ the volume form of the 4-cycle that $F_4$ is filling, whose directions (by the coordinates $z^i$), together with the $y$-direction, span the space transverse to the D4-brane. The $\delta$-function is the Dirac distribution that vanishes outside the D4-brane worldvolume.

The D4-brane induces a change of the $F_4$-flux on either side of the D4-brane, here for $y<0$ and $y>0$, modulo potential identifications. This can easily be seen when backreaction in the 4-cycle is suppressed (or effectively smearing the D4-branes over the 4-cycle). The Bianchi identity can then be solved for
\be
    F_4 = [N + Q_\text{D4} \theta (y)] \, \epsilon_4 \,,
\ee
where $\theta$ is the Heaviside step-function, and thus the flux undergoes a jump from one side to the other. For spherical domain walls in 4d spacetime, $y$ is the local radial coordinate and for the space-filling D4-branes $y$ is the local ``radial'' coordinate inside $\Sigma_2$, the 2-cycle Poincaré-dual to the 4-form $\epsilon_4$. 

\begin{figure}[h]
    \centering
    \begin{tikzpicture}[scale=0.38]
    
    \fill[color=red, opacity=0.1] (0,0) circle (5);
    \begin{scope}
        \clip (-5,-3) rectangle (5,5);
        \fill[color=orange, opacity=0.2] (0,0) circle (5);
    \end{scope}
    \fill[color=orange, opacity=0.2, even odd rule] (-4,-3) to[out=-18,in=198] (4,-3) (-4,-3) to (4,-3);
    
    \draw[thick, color=magenta, dotted] (4,-3) to[out=162,in=18] (-4,-3);
    \draw[thick, color=magenta] (-4,-3) to[out=-18,in=198] (4,-3);
    \draw[color=magenta, ->] (-2.6,-3.2) to (-2.6,-2.8);
    \draw[color=magenta, ->] (2.6,-3.2) to (2.6,-2.8);
    \draw[color=magenta, ->] (0,-3.6) to (0,-3.2);
    
    \node[] at (-5.5,3.5){$\Sigma_2$};
    \node[] at (0,-4.4){$N+1$};
    \node[] at (0,-0.8){$N$};
    \node[below right] at (4,-3){D4-brane};
    
    \begin{scope}[xshift=500pt]
        \draw[ultra thin, color=cyan, fill=white!90!cyan] (0,0) ellipse (6.5 and 5.0);
        \begin{scope}[xshift=80pt, yshift=50pt, xscale=0.8,yscale=0.6]
            \clip (0,2.2) ellipse (3 and 3.5);
            \draw[ultra thin, color=cyan, fill=white] (0,-2.2) ellipse (3 and 3.5);
        \end{scope}
        \begin{scope}[xshift=80pt, yshift=50pt, xscale=0.8,yscale=0.6]
            \clip(0,-1.8) ellipse (3 and 3.5);
            \draw[ultra thin, color=cyan] (0,2.2) ellipse (3 and 3.5);
        \end{scope}
        
        \begin{scope}[xshift=-90pt, yshift=-40pt, xscale=0.5,yscale=0.6]
            \clip (0,2.2) ellipse (3 and 3.5);
            \draw[ultra thin, color=cyan, fill=white] (0,-2.2) ellipse (3 and 3.5);
        \end{scope}
        \begin{scope}[xshift=-90pt, yshift=-40pt, xscale=0.5,yscale=0.6]
            \clip(0,-1.8) ellipse (3 and 3.5);
            \draw[ultra thin, color=cyan] (0,2.2) ellipse (3 and 3.5);
        \end{scope}
        
        \node[] at (2,-3.5){$F_4$};
        \node[] at (-6.5,3.5){$\Sigma_4$};
    \end{scope}
    
    \end{tikzpicture}
    \caption{A sketch of the D4-brane configuration that interpolates between $N$ and $N+1$ units of 4-form flux: the D4-brane (in magenta) wraps the AdS$_4$-spacetime and a contractible $1$-cycle that moves spanning a 2-cycle.} \label{fig:DGKT_D4-brane}
\end{figure}
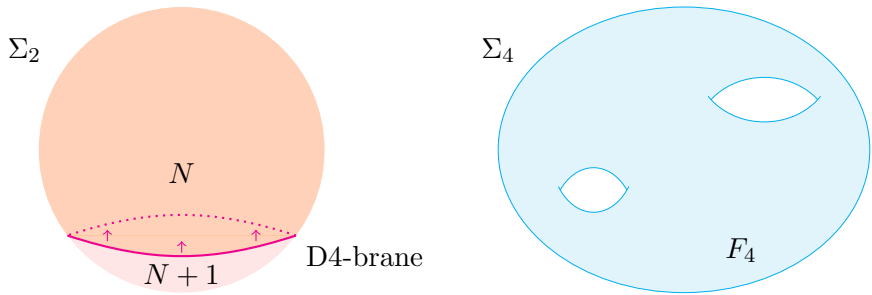

The fact that the $F_4$-flux differs on either side of the D4-brane is something we can utilise to our advantage. Indeed, one can use the creation and annihilation of such D4-branes to transition from one flux vacuum to the other. If we create such a D4-brane at a point in the 2-cycle, let it expand and contract again such that it sweeps out the whole 2-cycle, eventually annihilating against itself, it leaves a different flux vacuum behind. This is depicted in fig. \ref{fig:DGKT_D4-brane}.
In the DGKT construction, we have three different $F_4$-fluxes on three different 4-cycles. We therefore need to consider three such D4-branes to alter all fluxes.
The motion of these three D4-brane position moduli allows us to make an estimate of the distance between different flux vacua.

\subsection{Unwinding the flux for IIB vacua}
\begin{figure}[!h]
    \centering
    \begin{tikzpicture}[scale=0.24]

    \draw[thick, color=cyan, fill=gray!5!white] (-4,0) circle (5);
    \draw (0.25,1.0) rectangle (1.25,2.0);

    \begin{scope}
        \clip (9,-10) rectangle (29,10);
        
        \fill[color=gray!5!white] (-6,-11) circle (27);
        \draw[color=cyan] (-6,-11) circle (27);
        \draw[color=cyan] (-6,-11) circle (29);
        
        \draw[color=cyan, fill=cyan!10!white, even odd rule] (-6,-11) circle (27) (-6,-11) circle (29);
    \end{scope}
    
    \draw[ultra thin, gray] (0.25,1.0) to (9,-10);
    \draw[ultra thin, gray] (1.25,1.0) to (9,-1);
    \draw[ultra thin, gray] (1.25,2.0) to (9,3);
    \draw[ultra thin, gray] (0.25,2.0) to (9,10);
    \draw[thin, gray] (9,-10) rectangle (29,10);
    
    \draw[thin, dotted, ->] (23,2) to[out=270,in=20] (20.40,1);
    \draw (23,4) circle (2);
    \draw[color=magenta] (23-1.732,5) to[out=-15,in=195] (23+1.732,5) node[right,color=black]{NS5'};
    \draw[densely dotted, color=magenta] (23+1.732,5) to[out=165,in=15] (23-1.732,5);
    
    \draw[thin, dotted, ->] (15,-1) to[out=90,in=200] (18.66,0);
    \draw[fill=white] (15,-3) circle (2);
    \draw[color=magenta] (15-1.732,-4) node[left,color=black]{NS5'} to[out=-15,in=195] (15+1.732,-4);
    \draw[densely dotted, color=magenta] (15+1.732,-4) to[out=165,in=15] (15-1.732,-4);
    
    \node[align=center] at (-4,-0.5){$(K-1)_H$ \\ $(M-1)_{\text{D3}}$};
    \node[align=center] at (-14,2.5){$K_H$ \\ $1_{\overline{\text{D3}}}$};
    \node[] at (-4,-6){NS5 wall};
    
    \end{tikzpicture}
    \caption{There are two NS5-branes that move together due to the Freed-Witten effect. The NS5-brane domain wall in 4d spacetime (in cyan) wraps the 3-cycle. Inside and outside of it are $M-1$ D3-branes and $1$ anti-D3-brane. Those D3-branes puff into NS5'-branes (in magenta) that wrap a contractible 2-cycle inside $\Sigma_3$. Since the NS5'-branes are attached to the NS5-brane, the whole discharge process in 4d spacetime and inside the extra dimensions happens in a coordinated fashion.} \label{KPV}
\end{figure}
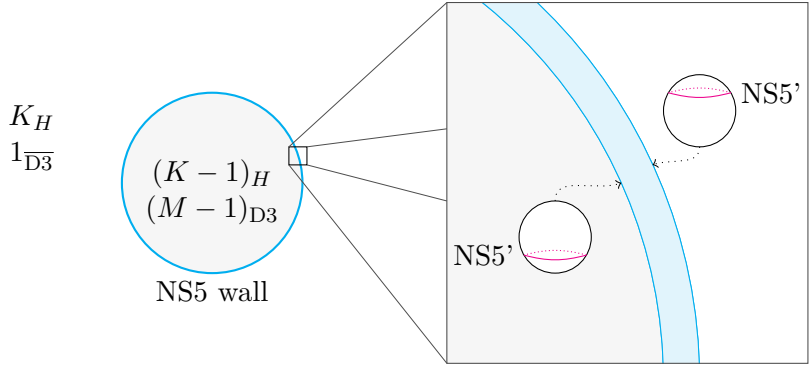

The interpolating brane setup in type IIA is reminiscent of what happens in type IIB vacua based on 3-form fluxes, such as GKP, KKLT and LVS. The 3-form fluxes can unwind with five-brane domain walls that act like Brown-Teitelboim bubbles in 4d spacetime. One can verify this in case the flux unwinding is dynamically triggered by a SUSY-breaking effect, namely anti-D3-branes. Inside a KS-like throat \cite{Klebanov:2000hb}, this is the familiar KPV process \cite{Kachru:2002gs}, which is also depicted in fig. \ref{KPV}.

In this case there is a decay of NSNS-flux from $K$ to $K-1$ units and at the same time $M$ D3-branes are created in the process that annihilate the original anti-D3-brane such that $M-1$ D3-branes are left at the end of the process. Here, $M$ is the RR-3-form flux on the dual 3-cycle. The creation of the D3-branes is necessary due to D3-tadpole cancellation. Now, this whole process can be viewed in two ways: one can say there is a spherical NS5-brane domain wall in 4d spacetime separating the two vacua. However, from the point of view of the anti-D3-brane, what happens is that the anti-D3-brane is really a spacetime-filling  NS5-brane (this is the NS5'-brane in fig. \ref{KPV}) with worldvolume flux that wraps a contractible cycle inside the 3-cycle $\Sigma_3$ wrapped by $F_3$-flux. So there are two NS5-branes here: the NS5-brane domain wall wrapping the whole 3-cycle $\Sigma_3$ and an NS5'-brane that acts like the 3-branes wrapping a contractible 2-cycle $\Sigma_2$ inside $\Sigma_3$. It is the dynamics of the position modulus of the NS5'-brane that was studied in detail in ref. \cite{Kachru:2002gs}. Both NS5-branes work together because of the Freed-Witten effect; an NS5-brane wrapping a cycle filled with $M$ units of $F_3$-flux has $M$ D3-branes attached to it.  These D3-branes really puff into the NS5'-branes. So the $H$-flux changes here happen both inside the 3-cycle and inside 4d space and the two discharges are identical.

In our 4d type IIA setting however, we do not have a Freed-Witten effect that connects the instanton and domain wall processes at the same time. Nevertheless, the D4-branes that we will use are reminiscent of the NS5'-branes discussed above, and our analysis below therefore will draw some parallels with the KPV process \cite{Kachru:2002gs}.

\subsection{A fantasy}
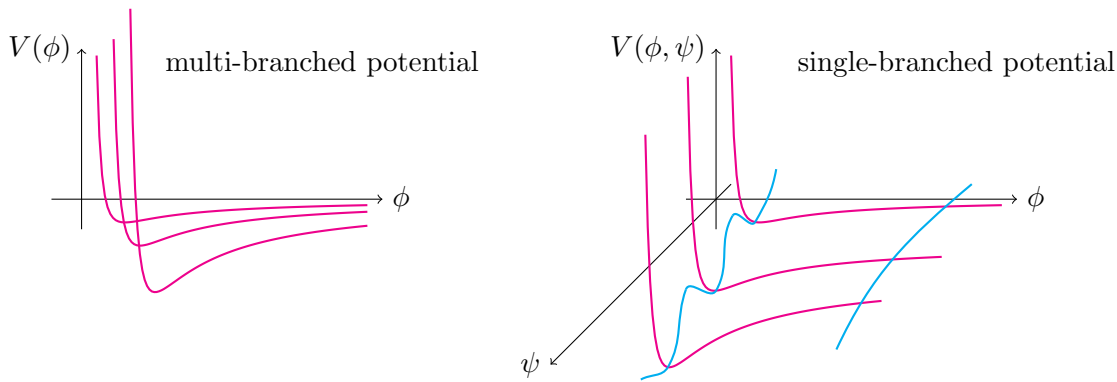
\begin{figure}[h]
    \centering
    \begin{tikzpicture}[scale=0.40]
    
    \begin{scope}[xshift=-300pt]
        
        \draw[->] (-1,0) -- (10,0) node[right]{$\phi$};
        \draw[->] (0,-1) -- (0,5) node[left]{$V(\phi)$};
        
        \draw[domain=0.5:9.5, thick, variable=\x, magenta, samples=100] plot ({\x}, {-2 + 2*exp(1/\x) - 4/\x});
        \draw[domain=1.065:9.5, thick, variable=\x, magenta, samples=100] plot ({\x}, {-4 + 4*exp(1/(\x-0.5)) - 8/(\x-0.5)});
        \draw[domain=1.62:9.5, thick, variable=\x, magenta, samples=100] plot ({\x}, {{-8 + 8*exp(1/(\x-1.0)) - 16/(\x-1.0)}});
        \node[] at (8,4.5){multi-branched potential};
        
    \end{scope}
    
    \begin{scope}[xshift=300pt]
        
        \draw[->] (-1,0) -- (10,0) node[right]{$\phi$};
        \draw[->] (0.5,0.5) -- (-5.5,-5.5) node[left]{$\psi$};
        \draw[->] (0,-1) -- (0,5) node[left]{$V(\phi,\psi)$};
        
        \draw[domain=0.5:9.5, thick, variable=\x, magenta, samples=100] plot ({\x}, {-2 + 2*exp(1/\x) - 4/\x});
        \draw[domain=-0.94:7.5, thick, variable=\x, magenta, samples=100] plot ({\x}, {-5.5 + 4*exp(1/(\x+1.5)) - 8/(\x+1.5)});
        \draw[domain=-2.35:5.5, thick, variable=\x, magenta, samples=100] plot ({\x}, {{-10.5 + 8*exp(1/(\x+3.0)) - 16/(\x+3.0)}});
        
        \draw[thick, cyan] (2,1) to[out=260,in=60] (1.3,-0.7607);
        \draw[thick, cyan] (1.3,-0.7607) to[out=240,in=60] (0.55,-0.55);
        \draw[thick, cyan] (0.55,-0.55) to[out=240,in=60] (0,-3.0424);
        \draw[thick, cyan] (0,-3.0424) to[out=240,in=70] (-1,-3);
        \draw[thick, cyan] (-1,-3) to[out=250,in=60] (-1.6,-5.5865);
        \draw[thick, cyan] (-1.6,-5.5865) to[out=240,in=30] (-2.5,-6.0);
        
        \draw[thick, cyan] (4,-5) to[out=65,in=220] (8.5,0.5);
        
        \node[] at (8,4.5){single-branched potential};
    \end{scope}
    
    \end{tikzpicture}
    \caption{A multi-branched potential (on the left) becomes single-branched (on the right) by integrating-in the open-string degree of freedom $\psi$.} \label{fig:FantasyPotential}
\end{figure}

Let us recall the conjecture that any two Calabi-Yau manifolds are connected to each other by a sequence of resolutions or deformations, known as Reid's fantasy \cite{reid1987moduli}. 
Resolutions and deformations are not smooth processes, resulting in topology changes. Yet one can visualise this connectedness by curves interpolating inside moduli space, deforming one Calabi-Yau into another one. Inspired by this and by our discussion above, we suggest an extension of this fantasy to moduli-stabilised scenarios, involving flux.
Once we have fluxes there are more integers than just the Betti-numbers (and other topological indices): we now have flux quanta as well that characterise the vacua. All these vacua can be thought of the minima of a multi-branched potential; see fig. \ref{fig:FantasyPotential}.
We suggest that we can interpolate between them through instanton processes that have a coarse-grained description in terms of scalar-field motion. Such scalars fields describe the position of space-filling branes moving on contractible cycles inside the compact geometry; see fig. \ref{fig:FantasyPotential}.
This is precisely what we suggested before in the context of the massive IIA vacua, where space-filling D4-branes achieve this. In particular, this means that, instead of considering different scalar potentials -- one for each discrete choice of fluxes --, we imagine to deal with a single-branched potential in which the different flux choices are connected through extra directions in the moduli space. In other words, we suggest that integrating-in open string fields turns the multi-branched flux potentials into single-branched potentials. Similar discussions of this integrating-in in the context of axion monodromy inflation \cite{Silverstein:2008sg,McAllister:2008hb,Marchesano:2014mla,Blumenhagen:2014gta,Hebecker:2014eua} can be found in ref. \cite{Brown:2016nqt}.

\section{Open-string scalar field and the distance conjecture} \label{sec: open-string SDC}
In what follows we are interested in computing the distance travelled in field space by D4-brane position-controlling scalars if we hop between different IIA orientifold flux vacua.

We start by reminding what the distance conjecture would entail for such IIA vacua: for super-Planckian field distances, there must appear in the spectrum a tower of states that become exponentially light. The mass scale of the tower $m$ shrinks with respect to a reference scale $m_0$ as
\begin{equation}
    \frac{m}{m_0} = \e^{- \beta \Delta},
\end{equation}
where $\Delta$ is the geodesic field displacement.
As anticipated above, in the IIA setup we can envision a new scalar field that changes the $F_4$-flux. Hence, a field displacement is changing the $F_4$-flux and therefore the masses of the fields will also change. We know that the masses scale with some power of the $F_4$-flux quantum $N$, since this controls all relevant moduli such as the dilaton and the volume. In other words, we expect a behaviour
\begin{equation}
    \frac{m}{m_0} \sim N^{-k},
\end{equation}
where $k$ is a constant that can be fixed unambiguously once the proper tower of light states has been identified. If the distance conjecture is to hold, the field displacement should behave as
\begin{equation}
\label{eq:SDC_DGKT}
    \Delta = \dfrac{k}{\beta} \log N\,.
\end{equation}

\subsection{D4-brane scalar-field distance}
The scalar field we wish to include in the effective description describes the position of the contractible D4-brane inside the compact geometry and we will denote it as $\psi$. A local description of the string-frame metric near the brane is
\begin{equation}
    \dd s_{10}^2 = g_{\mu\nu} \dd x^\mu \dd x^\nu +\ell_{s}^2 L^2 \left[f^2(\psi) \dd \varphi^2 + \dd \psi^2\right] + \de s_{\Sigma_4}^2.
\end{equation}
The D4-brane is extended along 4d spacetime -- which is an AdS$_4$-space with metric $g_{\mu\nu}$ -- wraps the circular direction $\varphi$ and is a point along the circle parametrised by $\psi$, where $\smash{\de s_{\Sigma_4}^2}$ is a metric in the leftover orthogonal space. Here, $L$ denotes the size of the 2-cycle $\Sigma_2$ in string units. The function $f(\psi)$ for a two-sphere is $f(\psi) = \sin \psi$, but we leave it here arbitrary as it in principle depends on the precise geometry of the 2-cycle. Let us now allow $\psi$ to vary with spacetime coordinates. Once pulled-back onto the brane, the metric then takes the form
\begin{equation}
    \dd s_{\text{D4}}^2 = \left(g_{\mu \nu} + \ell_{s}^2 L^2\partial_\mu \psi \partial_\nu \psi\right)\dd x^{\mu}\dd x^\nu + \ell_{s}^2 L^2 f^2(\psi) \dd \varphi^2. 
\end{equation}
When plugging this into the DBI-action of a single D4-brane, we get
\begin{equation}
    S_\text{D4} = - 2 \pi \ell_{s} L \mu_4 \int \dd^4 x \sqrt{-g_4} \; \e^{-\phi} |f(\psi)| \sqrt{1 + \ell_{s}^2 L^2 (\partial \psi)^2} \,,
\end{equation}
where $\mu_4 = 2 \pi / (2 \pi \ell_s)^5$ is the D4-brane tension. This, expanding the square root, gives
\begin{align}
    S_\text{D4}
    & = -2 \pi \ell_{s} \mu_4 \int \dd^4 x \sqrt{-g_4} \; \left[ \frac{1}{2} \ell_{s}^2 L^3 \e^{-\phi}|f(\psi)| (\partial \psi)^2 + L \, \e^{-\phi}|f(\psi)| \right],
\end{align}
which therefore gives a $\psi$-field moduli-space metric
\begin{equation}
    g_{\psi \psi} = 2\pi \mu_4 \ell_{s}^3 L^3 \e^{-\phi}|f(\psi)|.
\end{equation}
The idea is that, when the D4-brane reaches the poles, it pinches and leaves a vacuum behind with a change of $F_4$-flux on the dual 4-cycle by one unit. So we let $\psi = 0$ correspond to a configuration without $F_4$-flux and $\psi = \pi N$ to one with $N$ units. The discussion so far is only based on the DBI-action: this is sufficient to compute the kinetic term. As concerns the scalar potential, CS-actions give contributions of the same order as their DBI-counterparts (at least at supersymmetric points), so this analysis is enough to infer the relevant order of magnitude.

To compute the distance in scalar field space, one has to consider the ratio
\begin{equation}
    \frac{g_{\psi \psi}}{\Mpl^2} = \gamma \, \dfrac{\e^{\phi}}{L^3} |f(\psi)|,
\end{equation}
where, assuming that $L$ controls the size of the whole internal space, we used the identification $8 \pi^3 \gamma \Mpl^2 = L^6 \e^{-2\phi}\ell_{s}^{-2}$, for some order-one constant $\gamma$. We are interested in the distance in scalar field space between a vacuum with 1 unit of 4-form flux and one with $N$ units. For a theory of scalar fields $\varphi^\alpha$, when moving along a trajectory in the moduli space, parametrised by $s$, the distance in Planck units\footnote{Notice that, in general, the Planck mass has to be integrated, since it is a function of the moduli-space fields that are being integrated over.} is 
\begin{equation}
    \Delta = \int_{0}^{1} \dd s \, \frac{1}{\Mpl} \sqrt{g_{\alpha \beta} \, \dfrac{\de \varphi^\alpha}{\de s} \dfrac{\de \varphi^\beta}{\de s}}.
\end{equation}
In particular, the distance appearing in the SDC is computed along a geodesic path. Here, we can also exploit reparameterisation invariance in order to impose that $s=0$ represents the vacuum with $1$ unit of $F_4$-flux and $s=1$ to a vacuum with $N$ units of flux.

We should determine the distance by considering a geodesic in the field space.\footnote{If a non-geodesic path is used, one could only obtain an upper bound on the distance which might not be useful. For example, when we consider a path where we move the dilaton $\phi$ and volume scalar $L$ to their new vacuum values first and then move the brane modulus $\psi$, the distance will grow like $\Delta \sim N^{1/4}$. Using this upper bound would be too conservative, as we will see.} In order to do so correctly, we must treat the three D4-branes on the three different 2-cycles independently. It is then useful to work in terms of the 4d dilaton $D = \phi - \sum_{i=1}^3 \sigma_i$, since this field does not mix with the string-frame volumes $L_i = \e^{\sigma_i}$. A parametrisation of the string-frame metric is 
\begin{equation}
    \dd s_{10}^2 = \e^{2 D} \, \dd \tilde{s}_4^2 + \sum_{i=1}^3 \e^{2 \sigma_i} \, \dd s_{2,i}^2,
\end{equation}
where $\dd s_{2,i}^2$ represents the metrics of the 2-cycle $\Sigma_i$, and the 4-dimensional dilaton and the string-frame volumes indeed have a diagonal field-space metric with components $g_{DD} = 2 \Mpl^2$ and $g_{\sigma_i \sigma_i} = 2 \Mpl^2$. In view of this, a generic field-space distance can be written as
\begin{equation}
    \Delta = \int_0^1 \dd s \, \sqrt{\gamma \, \e^{D-\sum_{j=1}^3 \sigma_j} \!\! \sum_{i=1,2,3} \!\! \e^{3\sigma_i} |f(\psi_i)| \left(\frac{\de \psi_i}{\de s} \right)^2 + 2 \left(\frac{\de D}{\de s} \right)^2 + 2 \!\! \sum_{i=1,2,3} \!\! \left(\frac{\de \sigma_i}{\de s} \right)^2}.
\end{equation}
Because of the permutation symmetry in $i=1,2,3$, the geodesics will be such that the volume scalars of the three two-cycles are governed by the same solution, and hence we can work in terms of a single variable $\sigma = \sigma_1 = \sigma_2 = \sigma_3$. Hence, the geodesics for all D4-brane position scalars are the same and we can similarly just consider a single field $\psi = \psi_1 = \psi_2 = \psi_3$. In this case, the field-space distance becomes
\begin{equation}
    \Delta = \int_0^1 \dd s \, \sqrt{8 \, \e^{D} \left(\frac{\de x}{\de s} \right)^2 + 2 \left(\frac{\de D}{\de s} \right)^2 + 6 \left(\frac{\de \sigma}{\de s} \right)^2},
\end{equation}
where a new angular variable $x$ has also been defined for convenience, namely
\begin{equation}
    \frac{\de x}{\de \psi} = \sqrt{\dfrac{3 \gamma}{8}} \sqrt{|f(\psi)|}.
\end{equation}
Let us now define the variable $z = \e^{-D/2}$: the distance then becomes
\begin{equation}
\label{eq:distance_theta_z_w}
    \Delta = \int_0^1 \dd s \sqrt{\dfrac{8}{z^2} \left[\left(\frac{\de x}{\de s} \right)^2 + \left(\frac{\de z}{\de s} \right)^2 \right] + 6 \left(\frac{\de \sigma}{\de s} \right)^2}
\end{equation}
and the moduli space thus becomes a hyperbolic plane $\mathbb{H}^2$ spanned by $(x,z)$ and a line $\mathbb{R}$ spanned by $\sigma$.
The actual geodesic equations can be solved for 
\begin{equation}
    x(s) = l \sin [h(s)]+ x_0, \qquad
    z(s) = l \cos [h(s)], \qquad
    \sigma(s) = d_3 s + d_4,
\end{equation}
with the argument of the sine and cosine
\begin{equation}
    h(s) = 2 \arctan\left[\tanh \left(\dfrac{d_1 s + d_2}{2} \right)\right].
\end{equation}
All the parameters $d_i$, $l$ and $x_0$ are integration constants that are fixed by the boundary conditions of the geodesic curve. Note that the solutions to the geodesic equation can also be written down in terms of exponentials as
\begin{equation} \label{eq:geodesics}
    x = l \, \frac{e^{d_1 s + d_2}-1}{e^{2(d_1 s+ d_2)}+1} + x_0, \qquad
    z = 2l \, \frac{ \e^{d_1 s + d_2}}{e^{2(d_1 s+ d_2)}+1}, \qquad
    \sigma(s) = d_3 s + d_4.
\end{equation}
In both formulations, it is easy to see that the geodesics satisfy
\begin{equation}
    (x - x_0)^2+ z^2 = l^2,
\end{equation}
with $x_0$ and $l$ the same integration constants as before, and hence the geodesics are circular in the $(x,z)$-plane which is to be expected for hyperbolic spaces. This has now to be plugged back into \eqref{eq:distance_theta_z_w} to compute the shortest distance. The integral can be done explicitly, and one finds the very simple expression
\begin{equation}
\label{eq:distance}
    \Delta = \sqrt{8 d_1^2 + 6 d_3^2}.
\end{equation}
Hence, estimating the distance at large $N$ boils down to knowing the large-$N$ behaviour of the constants.
At $s=0$, we want the quantities to represent the vacuum with $N=1$, whereas we want $s=1$ to represent the vacuum with $N$ units of 4-form flux. We know that $\sigma = \log L $ and hence $\sigma(0) \sim 1$ and $\sigma(1)\sim \log N^{1/4}$ require $d_3 = \log N^{1/4}$ at leading order, as can be easily seen from eq. \eqref{eq:geodesics}. For the behaviour of $d_1$, one has to impose the boundary conditions on $x$ and $z$. We now have that $z = \e^{-D/2} = \e^{-\phi/2}L^{3/2}$, and hence we want $z(0)\sim 1$ and $z(1)\sim N^{3/4}$. The field $x$ represents our angle, and we want $x(0) \sim 1$ and $x(1) \sim N$. It is more difficult to see immediately from eq. \eqref{eq:geodesics} what $d_1$ should be, but one can confirm that it imposes $d_1^2 = (\log N^{5/4})^2$ at leading order.
With this we find from \eqref{eq:distance} the remarkable result that the distance scales with the flux quanta $N$ as
\begin{equation} \label{eq:finaldistance}
    \Delta \simeq  c \, \log N.
\end{equation}
Hence, the distance conjecture is satisfied due to eq. \eqref{eq:SDC_DGKT}! By direct inspection, we find the value $c = \sqrt{103/8}$. From standard worldsheet considerations, it is expected that the tower of states becoming exponentially light in the SDC is the KK-tower. This implies the constant $\beta$ appearing in the SDC to be $\beta = 1/\sqrt{206} \simeq 0.07$.

\subsection{Energy scales}
Concerning the discussion above, we have to check whether we can consistently think of $\psi$ as a scalar field that belongs to the EFT and understand the reliability of the conclusions that have been drawn based on the scalings in its effective action.

Usually, one would consider the field in its vacuum (in this case, this is $\psi=\pi \mathbb{Z}$), compute its mass and check whether the latter is below the cut-off. This is slightly problematic here since the potential is not necessarily differentiable in the vacuum. To see this, we can think of the case where $f(\psi)$ describes a 2-sphere, with the potential being
\begin{equation}
    V(\psi) = 2 \pi \ell_{s} \mu_4 \, L \, \e^{-\phi} |f(\psi)|.
\end{equation}
Clearly, $\psi=0$ is a vacuum since we know, by construction, that the D4-brane in its vacuum state pinches and annihilates itself, and yet $\alpha'$-corrections are expected to turn $V$ into a smooth function, whose details are beyond our knowledge and so the effective mass of this scalar particle; for a discussion about curvature corrections of an NS5-brane, see ref. \cite{Hebecker:2022zme} (building on earlier results on brane induced curvature terms \cite{Bachas:1999um, Junghans:2014zla}). Nonetheless, a heuristic analysis suggests that the mass of the field $\psi$ scales as a KK-mass. One can estimate this as follows: $\psi$-derivatives of the potential will not alter its scaling, implying the condition $\partial^2_{\psi} V(\psi) \sim V(\psi)$, which means that the effective canonically normalised mass scales like $m_\psi^2 \sim g_{\psi \psi}^{-1} V(\psi) \sim 1/L^2$.
So, such an effective mass is always at the order of the KK-scale,\footnote{Here a distinction can be made: the KK-scale for bulk modes is indeed $m_{\mathrm{KK}}^2 \simeq 1 / L^2$. However, the KK-mass for the D4-brane worldvolume fields is actually $m_{\mathrm{KK}}^2 \simeq 1 / [L^2 f^2(\psi)]$: for a 2-sphere, this becomes larger than the bulk KK-scale whenever the D4-brane is wrapped close to the poles.}
i.e. $m_\psi \sim m_{\mathrm{KK}} \sim N^{-1/4}$ in string units. This
is potentially dangerous since the KK-mass scale is the cutoff scale for the EFT.\footnote{It should be noted that this is again analogous to the KPV-construction \cite{Kachru:2002gs}. Here, far from the poles, the effective canonically-normalised scalar for the $\mathrm{S}^2$-sphere position modulus $\psi^{\mathrm{KPV}}$ can be seen to be of order $\smash{m^2_{\psi^{\mathrm{KPV}}} \simeq (m_{\mathrm{KK}}^{\mathrm{w}})^2}$, where $\smash{m_{\mathrm{KK}}^{\mathrm{w}}}$ is the warped KK-scale \cite{Burgess:2006mn}.}

To justify the previous claims on the distance conjecture being satisfied, a generic consideration on energy scales is helpful. We can compute the scale of the energy barrier in the D4-brane potential needed to hop to the other side and compare it with the AdS vacuum energy: the latter should clearly be an energy density well below the cut-off, by construction. On the one hand, the $\psi$-field scalar potential goes like $V(\psi) = 2\pi L \, \e^{-\phi}|f(\psi)| \sim N$. On the other hand, the pure DGKT flux potential, in string units, scales as $V_\text{flux} \sim N^{3/2}$. This means that the ratio of the potentials scales like
\begin{equation}
    \frac{V(\psi)}{V_\text{flux}} \sim N^{-1/2},
\end{equation}
and hence the energy density of the D4-brane will always be lower than the absolute value of the vacuum energy density. Note that this also means that the unstable vacua at the maximum of $f(\psi)$ are not de Sitter, consistent with expectations. As the scalar $\psi$ is also itself lighter than any string excitation, non-supersymmetric contributions from the heavier scalars in the string tower of states are less and less relevant as $N$ increases. We take these facts
as evidence that it is not problematic to include these D4-brane scalar fields into the DGKT effective action and apply the distance conjecture.

A more advanced consideration is the following. Because the KK-tower zero-mode has a mass comparable to the KK-scale, we must elaborate on what happens by including the heavier KK-states. By expanding the $\psi$-field as $\psi(x, \varphi) = \psi_0(x) + 2 \sum_{n =1}^\infty \psi_n(x) \cos{(n \varphi)}$, and adapting the analysis above, one finds highly non-trivial mixings of the zero-mode with all the other KK-modes, which should not be integrated out (at least, a non-zero finite number of them -- the lightest ones -- should not). Then, the whole analysis above would become a higher-dimensional problem, with the length that has been computed in the $(\psi, \phi, \sigma)$-space actually not being a geodesic distance. This, however, means that the actual geodesic in the $(\psi_0, \psi_1, \dots, \psi_\infty, \phi, \sigma)$-space will convey a shorter distance, being the actual geodesic. Indeed, the KK-states, which are fluctuations of the D4-brane position, are zero at the poles, and therefore the $(\psi, \phi, \sigma)$-space geodesic solves a constrained minimisation problem: in turn, this means that the full-space minimisation problem necessarily gives a distance that is smaller or equal than the former. Therefore, the $\log N$-behaviour found above can be considered as an upper bound, making the validation of the distance conjecture stronger.

However, a physical picture suggests that these mixings, although we argued can only improve the distance conjecture, might not be so essential. They are similar to fluctuations of Coleman-De Lucia instantons, which were found to be harmless \cite{Gratton:2000fj}, confirming the physical picture that symmetric bubbles are low-cost. In our current set-up the KK perturbations correspond to wiggles of the contractible circle inside the 2-cycles $\Sigma_2$. At the intuitive level one would not expect these wiggles to have any significant effect.

\subsection{Comments on non-Abelian D4-branes}
One could wonder whether the discussion above is fully general. It is not, since one may also consider a flux jump by $N$ units induced by the motion of a stack of $N$ D4-branes, as still suggested by the Bianchi identity. This corresponds to an immediate flux discharge/upcharge, as opposed to $N$ separate monodromies of Abelian branes.

If the D4-brane worldvolume is non-Abelian, the transverse scalars are in the adjoint representation of the $\mathrm{U}(N)$-group. Among these fall the scalars aligned in the $\psi$-direction: however, in this case the static gauge cannot be chosen such that $X^5 = \psi(\xi) L$. By expanding the field as $\psi = \psi_0 + \hat{\psi}$, where $\psi_0$ is the $\mathrm{U}(1)$-singlet and $\hat{\psi}$ is in the adjoint representation of the $\mathrm{SU}(N)$-subgroup, it seems natural to define the static gauge to be such that $X^5 = \psi_0(\xi) L$. In this case, expanding the non-Abelian D4-brane action \cite{Myers:1999ps}, one finds a kinetic term of the form $\mathrm{tr} \, g_{\psi \psi} \der \psi_0 \der \psi_0 = N g_{\psi \psi} \der \psi_0 \der \psi_0$. Similarly, the vacuum energy is also enhanced by a factor $\mathrm{tr} \, 1 = N$.

This has two consequences. First, a jump by $N$ units of flux is here associated to a metric that comes with an extra $N$-factor: this requires a change in the boundary conditions in the geodesic calculation, but it turns out not to alter the $\mathrm{log} \, N$-behaviour of the scalar-space distance. In particular, the constant $c$ in eq. \eqref{eq:finaldistance} happens to be smaller: this means that the distance is reduced slightly, as expected (since the flux upcharge happens in one time). Second, the vacuum energy is enhanced by a factor $N$, and therefore is actually parametrically larger than the DGKT-potential, being $V(\psi_0)/V_\text{flux} \sim \sqrt{N}$. So, although the immediate flux discharge would increase the scalar-field distance less than the equivalent chain of single-unit discharges, the energy that is associated to this configuration is too consider it as part of the EFT.

\section{Extensions to other classes of flux vacua} \label{sec: other classes of vacua}
Here we are going to explore the applicability of the ideas presented above for two more classes of flux vacua. First, we will discuss scale-separated massive type IIA AdS$_3$ vacua. Second, we will comment on the obstacles encountered for generic Freund-Rubin-type compacifications.

\subsection{D-brane motion in scale-separated AdS\texorpdfstring{$\boldsymbol{_3}$}{$_3$} vacua}\label{subsec: AdS3-vacua}
It has been shown in refs. \cite{Farakos:2020phe, VanHemelryck:2022ynr} that scale-separated AdS$_3$ vacua can also arise from massive type IIA orientifolds. There, the type IIA theory is compacified on a $G_2$-manifold with O2- and O6-planes. This can be seen as the 3d analogue of DGKT. For a SUSY solution (with smeared sources), the RR-fluxes are
\begin{subequations}
\begin{align}
    & g_s F_0 = -4 \mu,\\
    & g_s F_2 = 0, \\
    & g_s F_4 = - \star H - 2 \mu \star \Phi, \\    
    & g_s F_6 = 0,
\end{align}
\end{subequations}
where $\Phi$ is the left-invariant 3-form of $G_2$ and $\mu = L_\text{AdS}^{-1}$, and where the NSNS-3-form satisfies
\begin{equation}
    H \wedge \Phi  = 0, \qquad
    H \wedge \star \Phi  = -6 \mu \, \mathrm{vol}_7.
\end{equation}
The fluxes also have to satisfy the Bianchi identities. There are only O2- and O6-planes, and hence the only non-trivial Bianchi identities are those for $F_2$ and $F_6$, which read
\begin{equation}
    0  = \int H F_0 + \int j_\text{O6}, \qquad
    0  = \int H \wedge F_4 + \int j_\text{O2/D2}.
\end{equation}
We also included D2-branes in the compactification, as it turns out that scale separation is only attainable when the charge of the O2-planes cancel those of the D2-branes \cite{Farakos:2020phe,VanHemelryck:2022ynr}, which means that $\int H\wedge F_4 = 0$ for supersymmetric scale-separated vacua.
One could realise this setup on a $\mathrm{T}^7/ \mathbb{Z}_2 \times \mathbb{Z}_2 \times \mathbb{Z}_2$-orbifold. There are seven 3-cycles $\Sigma_{3,i}$ and seven dual 4-cycles $\Sigma_{4,i}$, with $i=1,...,7$.

We can then decompose the forms as
\begin{equation}
    \int_{\Sigma_{3,i}}\Phi = \ell_s^3 \e^{3\sigma^i}, \qquad
    \int_{\Sigma_{3,i}} H = (2\pi \ell_s)^2 h, \qquad
    \int_{\Sigma_{4,i}} F_4 = (2\pi \ell_s)^3 f \tilde{N}_i,
\end{equation}
where the $\tilde{N}_i$, $h$ and $f$ are integers. Additionally we require $\gcd(\tilde{N}_1,...,\tilde{N}_7)=1$ as a non-unity gcd can always be absorbed into the definition of the integer $f$. The $H$-flux on every 3-cycle is indeed the same as dictated by the $F_2$-Bianchi identity \cite{Farakos:2020phe}. 
The condition that $\int H \wedge F_4 =0$ then translates to the constraint
\begin{equation}
    \sum_{i=1}^{7} \tilde{N}_i = 0.
\end{equation}
Scale separation can be achieved by letting $f \gg 1$ with the following scalings in that limit:
\begin{equation}
    \e^{\sigma_i} \sim f^{1/4}, \qquad
    g_s \sim f^{-3/4}, \qquad
    \mu \sim f^{-3/4}.
\end{equation}
The 7d volume is given by $\smash{V_7 = \exp({\sum_{i=1}^7 \sigma_i})}$ and scales like $f^{7/4}$.
Scale separation is indeed achieved in the large $f$-limit as
\begin{equation}
    \frac{L_\text{KK}^2}{L_\text{AdS}^2} = \mu^2 V_7^{2/7} \sim f^{-1}.
\end{equation}
Similarly to the AdS$_4$ case above, one can induce a change in the $F_4$-flux on a certain 4-cycle by wrapping a D4-brane on a contractible 2-cycle inside the 3-cycle that is dual to that 4-cycle of the flux. One would expect this to go precisely as in the 4d-case, but there is an additional difficulty: all the 3-cycles have $H$-flux on them. It means that the pullback of the $B$-field on the D4-world-volume is not vanishing and induces other brane charges (here D2-charge), similarly as we are used to for KPV \cite{Kachru:2002gs}. Taking this into account, the distance between these AdS$_3$ flux vacua can analogously be derived. We refer the reader to appendix \ref{app:AdS3} for more details about the D4-brane setup for these AdS$_3$ vacua.

\subsection{From IIA orientifolds to  M-theory Freund-Rubin vacua} \label{subsec: Freund-Rubin vacua}

So far we have discussed transitions between vacua for scale separated AdS vacua of massive type IIA string theory with O6-planes, such as those pioneered in \cite{DeWolfe:2005uu} and \cite{Farakos:2020phe}. However, there are of course many other flux vacua that can also be classified by flux quanta. The simplest among such solutions are Freund-Rubin vacua, i.e. simple compactifications to AdS spacetimes on a positive Einstein manifold with flux either filling AdS or the internal manifold. The most famous examples are AdS$_5 \times \mathrm{S}^5$ with $F_5$-flux in type IIB string theory and AdS$_4 \times \mathrm{S}^7$ and AdS$_7 \times \mathrm{S}^4$ in M-theory with $G_7$- and $G_4$-flux.
Unfortunately, it is not straightforward to apply the same logic as before to transition between different flux vacua. Indeed, in AdS$_5 \times \mathrm{S}^5$ the branes that change the $F_5$-flux are D3-branes, but there is no cycle in which they can be codimension one and spacetime filling at the same time. Similarly, one cannot wrap M2- or M5-branes appropriatly in AdS$_4 \times \mathrm{S}^7$ and AdS$_7 \times \mathrm{S}^4$ to change the right flux. It would be interesting to identify the more exotic objects that do, and we leave that for future research.

The problem seems to appear whenever there is an electric flux, i.e. when it fills spacetime, which is equivalent to a flux filling the internal dimensions through Hodge duality. Other vacua with this feature are the vacuum solutions with O6-planes for massless IIA strings \cite{Cribiori:2021djm}, which are related through T-duality to the massive IIA models discussed above. These are not Freund-Rubin solutions, but their M-theory lifts appear to be so.
The massless IIA models have no $F_0$-, $F_4$- and $H$-flux, but do instead have non-vanishing $F_2$-, $F_6$- and geometric fluxes. Of the three $F_2$-fluxes, two are not bounded by tadpoles. The same is true for $F_6$, filling the whole internal manifold. The branes that would interpolate between different flux vacua should be two D6-branes of codimension one within the 4-cycles dual to the 2-cycles that these two $F_2$-fluxes fill. Such D6-branes are predicted by two T-dualities of the D4-brane setup in massive type IIA discussed before. 
For the $F_6$-flux it is a D2-brane that changes the flux, but this can only be a proper domain wall in 4d spacetime. However, T-duality of the D4-brane setup in massive type IIA suggests that one of the D4-branes there gets mapped to another D4-brane in the massless type IIA setting. With the right world-volume flux it would be able to carry D2-charge, and hence change the flux when the D4-brane moves. But it must do so without creating $F_4$-flux. The 6d manifold has torsional cycles and hence leaves indeed the possibility open that this D4-brane does not carry its own D4-charge \cite{Cascales:2004qp,Marchesano:2006ns}.
It is important to investigate whether this intuition is correct or whether more exotic objects are needed to interpolate between these massless type IIA vacua. Once understood, its M-theory uplift serves as a gateway to the exploration of the landscape of more general Freund-Rubin vacua. The presence of D6/D4 boundstates with D4's having worldvolume fluxes, sourcing D2-charges, in massless IIA suggests, after uplift that bound states of KK branes potentially with M- (or D-)branes with worldvolume fluxes are the sought-for transitions for Freund-Rubin vacua. We leave this for future research. 

\section{Conclusions} \label{sec: conclusions}

In this paper, we explained how a scalar field can interpolate between different vacua of massive type IIA orientifolds in which scale separation can be achieved. Standard practice is to use D4-domain walls to interpolate between the vacua, but, in this work, we used spacetime-filling D4-branes wrapped on contractible 1-cycles within 2-cycles to change the $F_4$-flux. In such a setup, the position moduli of the D4-branes determine the amount of $F_4$-flux and hence allow us to travel through the flux vacuum landscape.  This is reminiscent of a flux analogue of Reid's fantasy where different flux vacua are connected in field space. Next, we verified that these D4-brane states can be kept in the EFT and calculated the distance between the different flux vacua in this new field space. Quite surprisingly, we find that the swampland distance conjecture is satisfied for the scale separated IIA vacua. This is the main result of this paper and suggests that scale-separated vacua may not be in the swampland, in contradiction to the strong ADC conjecture \cite{Lust:2019zwm}. The latter conjecture was based on the distances travelled by the spacetime metric, which by itself is less well-motivated than the distance conjecture for scalar fields. Additionally, we argued that massive IIA 3d scale-separated vacua satisfy similar properties. To support our arguments, we made use of the ordinary brane actions and ignored curvature corrections of the D4-branes (along the lines considered in refs. \cite{Bachas:1999um, Junghans:2014zla}). Nevertheless, it would be interesting to investigate their effect in the same vein as ref. \cite{Hebecker:2022zme}.

Ideally, these ideas get extended in scope to describe flux transitions for Freund-Rubin vacua in type II string theory and M-theory, although a clear picture of how that occurs is still lacking. Nonetheless, a detailed M-theory lift of the vacuum transitions in the massless IIA scale separated vacua of ref. \cite{Cribiori:2021djm} should be possible since they can formally be related to the setup in this paper via double T-duality \cite{Caviezel:2008ik}. A subsequent uplift to 11-dimensional supergravity, where one should find scale separated Freund-Rubin vacua, should then clarify how vacuum transitions occur. 

So far, we have only discussed classical type IIA vacua. One could also consider the quantum vacua in type IIB. Scale separation is achieved there if the on-shell flux superpotential is exponentially small. The latter is determined by the RR- and NSNS-3-form fluxes and we see a potential avenue to change these fluxes by D5- and NS5-branes, as is done in the KPV-construction. However, these fluxes are bound by tadpoles. It would be interesting to see whether there exists a (finite) chain of vacua which are similarly connected by D5- and/or NS5-branes which wrap contractible 3-cycles and see whether the distance conjecture is satisfied.

\section*{Acknowledgements}
We would like to thank Fridrik Gautason, Fernando Marchesano and Miguel Montero for useful discussions. 
The work of GS and FT is supported in part by the DOE grant DE-SC0017647.
VVH would like to thank the University of Wisconsin-Madison for its hospitality while this work was performed. VVH is supported by research grant nr. 1185120N and travel grant V427822N of the Research Foundation - Flanders (FWO) and partially by the KU Leuven C1 grant ZKD1118 C16/16/005.

\appendix

\section{Space-filling D4-branes in scale-separated AdS\texorpdfstring{$\boldsymbol{_3}$}{$_3$} vacua}
\label{app:AdS3}
In this appendix we elaborate further on the precise D4-brane setup one should use to interpolate between different AdS$_3$ flux vacua. 
Let us take a look at $\tilde{N}_i$ D4-branes wrapping a 2-cycle within the 3-cycle $\Sigma_{3,i}$. That 3-cycle has volume $s_i$ and locally we parametrise the metric by
\begin{equation}
    \dd s_{10}^2 = \dd s_{\text{AdS}_3}^2 + \ell_s^2 \e^{2\sigma_i} \left[\dd \psi_i^2 + w_i^2(\psi_i) g_{mn} \dd y^m \dd y^n\right] + \de s_{\Sigma_{4,i}}^2,
\end{equation}
where the D4-brane is parallel to $y^m,y^n$, while $\psi_i$ is the transverse coordinate within the 3-cycle; moreover, $\Sigma_{4,i}$ is an extra transverse 4-cycle. Taking into account only the collective mode, the full brane action becomes
\begin{equation}
    S_{\text{D4},i} = -|\tilde{N}_i|\mu_4 \int \dd^5 x \sqrt{- \det(g+ 2\pi \ell_s^2 \mathcal{F})} + |\tilde{N}_i|\mu_4 \int C \wedge \e^{\mathcal{F}}.
\end{equation}
Here we need $\int_{\Sigma_{3,i}} \dd B_i = h$, which can be achieved by introducing a certain function $U_i(\psi_i)$ such that
\begin{equation}
    B_i = (2\pi\ell_s)^2 U_i(\psi_i) \tilde{\mathrm{vol}}(\Sigma_{2,i}), 
\end{equation}
where $\tilde{\mathrm{vol}}(\Sigma_{2,i})$ refers to the volume form of the contractible 2-cycle $\Sigma_i$ without the scale $\e^{2\sigma_i}$. So $\mathcal{F}_i$ is non-vanishing and we have $2\pi \ell_s^2 \mathcal{F}_i = B_i$. One can again change the $F_4$-flux on the dual 4-cycle by letting the brane go from one pole of the 3-cycle to the other, i.e. travelling in $\psi_i$. Doing so, a single brane generates $h$ units of D2-charge after traveling from one pole to the other. 

As discussed before, the $F_4$-fluxes have to sum to zero and no net D2-charge can appear for SUSY scale-separated vacua. This means that if we want to hop from one vacuum to another, increasing $f$, some $F_4$-fluxes need to decrease (the negative ones), while others need to increase. This means that not all branes need to be oriented in the same way. We could equivalently say that the D4-branes  associated to positive flux number $\tilde{N}_i$ travel in the positive $\psi_i$-direction, while the branes associated to negative $\tilde{N}_i$ travel along negative $\psi_i$. Hence, the vacua will be lying at the points $\psi_i = \text{sgn}(\tilde{N}_i) f h$. For the D2-charge, this means that for arbitrary $f$ we have
\begin{equation}
    U_i (\text{sgn}(\tilde{N_i}) \pi f) = \text{sgn}(\tilde{N_i})  f h.
\end{equation}

The resulting D4-brane action then becomes familiar as it looks very much like NS5-brane action of KPV, and with $v_{2,i} = \int \tilde{\mathrm{vol}}(\Sigma_{2,i})$ we find
\begin{equation}
    \begin{split}
        S_{\text{D4},i} = -|\tilde{N}_i| \, v_{2,i} \mu_4 \ell_s^2 \int \dd^3 x \sqrt{-g_4} \; \e^{2\sigma_i-\phi} \sqrt{w_i^4(\psi_i) + U_i^2(\psi_i)} \sqrt{1+ \e^{2\sigma_i} (\partial \psi_i)^2} \\
        + |\tilde{N}_i| \, \mu_3 v_{2,i} \, U_i(\psi_i) \int C_3 + |\tilde{N}_i| \, \mu_4 \int C_5 & .
    \end{split}
\end{equation}
If one considers all D4-branes together and assumes all the 3-cycles to have the same topology, so that $v_{2,i} = v_2 $ for each $i$, we get that the effective D2-charge is
\begin{equation}
    Q_\text{D2} = \mu_3 v_2 \sum_i |\tilde{N}_i| U_i (\text{sgn}(\tilde{N_i}) \pi f) = \mu_3 v_2 f h\sum_i \tilde{N}_i = 0.
\end{equation}
After expanding the square root with the kinetic term, one can now see that field space metric of the position scalar $\psi_i$ is given by
\begin{equation}
    g_{\psi_i \psi_i} = | \tilde{N}_i |\mu_4 v_2 \e^{4\sigma_i-\phi}.
\end{equation}
Given the 3d Planck mass $\Mpl = V_7 \, \e^{-2\phi} = \e^{- 2\phi + \sum_{i=1}^7 \sigma_i}$, we easily compute the ratio of the field space metric with $\Mpl$ to be
\begin{equation}
    \frac{g_{\psi_i \psi_i}}{\Mpl} =  \gamma | \tilde{N}_i | v_2\sqrt{w_i^4(\psi_i) + U_i^2(\psi_i)} \e^{\phi + 4 \sigma_i - \sum_{i=j}^7 \sigma_j}
\end{equation}
where $\gamma$ is again a constant.
This scales as $f^{3/2}$, just as its 4d analogue. Moreover, we find the potential 
\begin{equation}
    V(\psi_i) \supset | \tilde{N}_i| v_2\mu_4 \ell_s^2 \, \e^{2\sigma_i-\phi} \sqrt{w_i^4(\psi_i) + U_i^2(\psi_i)}
\end{equation}
which scales as $f^{5/4}$ in string units. The flux potential scales as $f^{7/4}$ and hence we have
\begin{equation}
    \frac{V(\psi_i)}{V_\text{flux}} \sim f^{-1/2},
\end{equation}
which is also the same as in the 4d case. A computation of the scalar field distance between different vacua (labelled by $f$) can then be done in an the same way as in the 4d case.  

\bibliographystyle{JHEP}
\bibliography{references.bib}

\providecommand{\href}[2]{#2}\begingroup\raggedright\begin{thebibliography}{10}

\bibitem{Derendinger:2004jn}
J.-P. Derendinger, C.~Kounnas, P.~M. Petropoulos and F.~Zwirner,
  \emph{{Superpotentials in IIA compactifications with general fluxes}},
  \href{https://doi.org/10.1016/j.nuclphysb.2005.02.038}{\emph{Nucl. Phys. B}
  {\bfseries 715} (2005) 211}
  [\href{https://arxiv.org/abs/hep-th/0411276}{{\ttfamily hep-th/0411276}}].

\bibitem{DeWolfe:2005uu}
O.~DeWolfe, A.~Giryavets, S.~Kachru and W.~Taylor, \emph{{Type IIA moduli
  stabilization}},
  \href{https://doi.org/10.1088/1126-6708/2005/07/066}{\emph{JHEP} {\bfseries
  07} (2005) 066} [\href{https://arxiv.org/abs/hep-th/0505160}{{\ttfamily
  hep-th/0505160}}].

\bibitem{Camara:2005dc}
P.~G. Camara, A.~Font and L.~E. Ibanez, \emph{{Fluxes, moduli fixing and
  MSSM-like vacua in a simple IIA orientifold}},
  \href{https://doi.org/10.1088/1126-6708/2005/09/013}{\emph{JHEP} {\bfseries
  09} (2005) 013} [\href{https://arxiv.org/abs/hep-th/0506066}{{\ttfamily
  hep-th/0506066}}].

\bibitem{Cribiori:2021djm}
N.~Cribiori, D.~Junghans, V.~Van~Hemelryck, T.~Van~Riet and T.~Wrase,
  \emph{{Scale-separated AdS4 vacua of IIA orientifolds and M-theory}},
  \href{https://doi.org/10.1103/PhysRevD.104.126014}{\emph{Phys. Rev. D}
  {\bfseries 104} (2021) 126014}
  [\href{https://arxiv.org/abs/2107.00019}{{\ttfamily 2107.00019}}].

\bibitem{Farakos:2020phe}
F.~Farakos, G.~Tringas and T.~Van~Riet, \emph{{No-scale and scale-separated
  flux vacua from IIA on G2 orientifolds}},
  \href{https://doi.org/10.1140/epjc/s10052-020-8247-5}{\emph{Eur. Phys. J. C}
  {\bfseries 80} (2020) 659}
  [\href{https://arxiv.org/abs/2005.05246}{{\ttfamily 2005.05246}}].

\bibitem{VanHemelryck:2022ynr}
V.~Van~Hemelryck, \emph{{Scale-separated AdS$_3$ vacua from $G_2$-orientifolds
  using bispinors}},  \href{https://arxiv.org/abs/2207.14311}{{\ttfamily
  2207.14311}}.

\bibitem{Kachru:2003aw}
S.~Kachru, R.~Kallosh, A.~D. Linde and S.~P. Trivedi, \emph{{De Sitter vacua in
  string theory}},
  \href{https://doi.org/10.1103/PhysRevD.68.046005}{\emph{Phys. Rev. D}
  {\bfseries 68} (2003) 046005}
  [\href{https://arxiv.org/abs/hep-th/0301240}{{\ttfamily hep-th/0301240}}].

\bibitem{Balasubramanian:2005zx}
V.~Balasubramanian, P.~Berglund, J.~P. Conlon and F.~Quevedo,
  \emph{{Systematics of moduli stabilisation in Calabi-Yau flux
  compactifications}},
  \href{https://doi.org/10.1088/1126-6708/2005/03/007}{\emph{JHEP} {\bfseries
  03} (2005) 007} [\href{https://arxiv.org/abs/hep-th/0502058}{{\ttfamily
  hep-th/0502058}}].

\bibitem{Lust:2019zwm}
D.~L\"ust, E.~Palti and C.~Vafa, \emph{{AdS and the Swampland}},
  \href{https://doi.org/10.1016/j.physletb.2019.134867}{\emph{Phys. Lett. B}
  {\bfseries 797} (2019) 134867}
  [\href{https://arxiv.org/abs/1906.05225}{{\ttfamily 1906.05225}}].

\bibitem{Collins:2022nux}
T.~C. Collins, D.~Jafferis, C.~Vafa, K.~Xu and S.-T. Yau, \emph{{On Upper
  Bounds in Dimension Gaps of CFT's}},
  \href{https://arxiv.org/abs/2201.03660}{{\ttfamily 2201.03660}}.

\bibitem{Lust:2022lfc}
S.~L\"ust, C.~Vafa, M.~Wiesner and K.~Xu, \emph{{Holography and the KKLT
  scenario}}, \href{https://doi.org/10.1007/JHEP10(2022)188}{\emph{JHEP}
  {\bfseries 10} (2022) 188}
  [\href{https://arxiv.org/abs/2204.07171}{{\ttfamily 2204.07171}}].

\bibitem{Cribiori:2022trc}
N.~Cribiori and G.~Dall'Agata, \emph{{Weak gravity versus scale separation}},
  \href{https://doi.org/10.1007/JHEP06(2022)006}{\emph{JHEP} {\bfseries 06}
  (2022) 006} [\href{https://arxiv.org/abs/2203.05559}{{\ttfamily
  2203.05559}}].

\bibitem{Montero:2022ghl}
M.~Montero, M.~Rocek and C.~Vafa, \emph{{Pure Supersymmetric AdS and the
  Swampland}},  \href{https://arxiv.org/abs/2212.01697}{{\ttfamily
  2212.01697}}.

\bibitem{Montero:2022prj}
M.~Montero, C.~Vafa and I.~Valenzuela, \emph{{The Dark Dimension and the
  Swampland}},  \href{https://arxiv.org/abs/2205.12293}{{\ttfamily
  2205.12293}}.

\bibitem{Polchinski:2009ch}
J.~Polchinski and E.~Silverstein, \emph{{Dual Purpose Landscaping Tools: Small
  Extra Dimensions in AdS/CFT}}, pp.~365--390.
\newblock 8, 2009.
\newblock \href{https://arxiv.org/abs/0908.0756}{{\ttfamily 0908.0756}}.
\newblock 10.1142/9789814412551\textunderscore0018.

\bibitem{Alday:2019qrf}
L.~F. Alday and E.~Perlmutter, \emph{{Growing Extra Dimensions in AdS/CFT}},
  \href{https://doi.org/10.1007/JHEP08(2019)084}{\emph{JHEP} {\bfseries 08}
  (2019) 084} [\href{https://arxiv.org/abs/1906.01477}{{\ttfamily
  1906.01477}}].

\bibitem{Aharony:2008wz}
O.~Aharony, Y.~E. Antebi and M.~Berkooz, \emph{{On the Conformal Field Theory
  Duals of type IIA AdS(4) Flux Compactifications}},
  \href{https://doi.org/10.1088/1126-6708/2008/02/093}{\emph{JHEP} {\bfseries
  02} (2008) 093} [\href{https://arxiv.org/abs/0801.3326}{{\ttfamily
  0801.3326}}].

\bibitem{Conlon:2018vov}
J.~P. Conlon and F.~Quevedo, \emph{{Putting the Boot into the Swampland}},
  \href{https://doi.org/10.1007/JHEP03(2019)005}{\emph{JHEP} {\bfseries 03}
  (2019) 005} [\href{https://arxiv.org/abs/1811.06276}{{\ttfamily
  1811.06276}}].

\bibitem{Conlon:2020wmc}
J.~P. Conlon and F.~Revello, \emph{{Moduli Stabilisation and the Holographic
  Swampland}}, \href{https://doi.org/10.31526/lhep.2020.171}{\emph{LHEP}
  {\bfseries 2020} (2020) 171}
  [\href{https://arxiv.org/abs/2006.01021}{{\ttfamily 2006.01021}}].

\bibitem{Conlon:2021cjk}
J.~P. Conlon, S.~Ning and F.~Revello, \emph{{Exploring the holographic
  Swampland}}, \href{https://doi.org/10.1007/JHEP04(2022)117}{\emph{JHEP}
  {\bfseries 04} (2022) 117}
  [\href{https://arxiv.org/abs/2110.06245}{{\ttfamily 2110.06245}}].

\bibitem{Apers:2022tfm}
F.~Apers, J.~P. Conlon, S.~Ning and F.~Revello, \emph{{Integer conformal
  dimensions for type IIa flux vacua}},
  \href{https://doi.org/10.1103/PhysRevD.105.106029}{\emph{Phys. Rev. D}
  {\bfseries 105} (2022) 106029}
  [\href{https://arxiv.org/abs/2202.09330}{{\ttfamily 2202.09330}}].

\bibitem{Apers:2022vfp}
F.~Apers, \emph{{Aspects of AdS flux vacua with integer conformal dimensions}},
   \href{https://arxiv.org/abs/2211.04187}{{\ttfamily 2211.04187}}.

\bibitem{Apers:2022zjx}
F.~Apers, M.~Montero, T.~Van~Riet and T.~Wrase, \emph{{Comments on classical
  AdS flux vacua with scale separation}},
  \href{https://doi.org/10.1007/JHEP05(2022)167}{\emph{JHEP} {\bfseries 05}
  (2022) 167} [\href{https://arxiv.org/abs/2202.00682}{{\ttfamily
  2202.00682}}].

\bibitem{Quirant:2022fpn}
J.~Quirant, \emph{{Noninteger conformal dimensions for type IIA flux vacua}},
  \href{https://doi.org/10.1103/PhysRevD.106.066017}{\emph{Phys. Rev. D}
  {\bfseries 106} (2022) 066017}
  [\href{https://arxiv.org/abs/2204.00014}{{\ttfamily 2204.00014}}].

\bibitem{Plauschinn:2022ztd}
E.~Plauschinn, \emph{{Mass spectrum of type IIB flux compactifications -
  comments on AdS vacua and conformal dimensions}},
  \href{https://arxiv.org/abs/2210.04528}{{\ttfamily 2210.04528}}.

\bibitem{Danielsson:2018ztv}
U.~H. Danielsson and T.~Van~Riet, \emph{{What if string theory has no de Sitter
  vacua?}}, \href{https://doi.org/10.1142/S0218271818300070}{\emph{Int. J. Mod.
  Phys. D} {\bfseries 27} (2018) 1830007}
  [\href{https://arxiv.org/abs/1804.01120}{{\ttfamily 1804.01120}}].

\bibitem{Gao:2020xqh}
X.~Gao, A.~Hebecker and D.~Junghans, \emph{{Control issues of KKLT}},
  \href{https://doi.org/10.1002/prop.202000089}{\emph{Fortsch. Phys.}
  {\bfseries 68} (2020) 2000089}
  [\href{https://arxiv.org/abs/2009.03914}{{\ttfamily 2009.03914}}].

\bibitem{Bena:2020xrh}
I.~Bena, J.~Bl\r{a}b\"ack, M.~Gra\~na and S.~L\"ust, \emph{{The tadpole
  problem}}, \href{https://doi.org/10.1007/JHEP11(2021)223}{\emph{JHEP}
  {\bfseries 11} (2021) 223}
  [\href{https://arxiv.org/abs/2010.10519}{{\ttfamily 2010.10519}}].

\bibitem{Emelin:2020buq}
M.~Emelin, \emph{{Effective Theories as Truncated Trans-Series and Scale
  Separated Compactifications}},
  \href{https://doi.org/10.1007/JHEP11(2020)144}{\emph{JHEP} {\bfseries 11}
  (2020) 144} [\href{https://arxiv.org/abs/2005.11421}{{\ttfamily
  2005.11421}}].

\bibitem{Junghans:2022exo}
D.~Junghans, \emph{{LVS de Sitter Vacua are probably in the Swampland}},
  \href{https://arxiv.org/abs/2201.03572}{{\ttfamily 2201.03572}}.

\bibitem{DallAgata:2022abm}
G.~Dall'Agata, M.~Emelin, F.~Farakos and M.~Morittu, \emph{{Anti-brane uplift
  instability from goldstino condensation}},
  \href{https://doi.org/10.1007/JHEP08(2022)005}{\emph{JHEP} {\bfseries 08}
  (2022) 005} [\href{https://arxiv.org/abs/2203.12636}{{\ttfamily
  2203.12636}}].

\bibitem{Farakos:2020wfc}
F.~Farakos, A.~Kehagias and N.~Liatsos, \emph{{de Sitter decay through
  goldstino evaporation}},
  \href{https://doi.org/10.1007/JHEP02(2021)186}{\emph{JHEP} {\bfseries 02}
  (2021) 186} [\href{https://arxiv.org/abs/2009.03335}{{\ttfamily
  2009.03335}}].

\bibitem{Junghans:2022kxg}
D.~Junghans, \emph{{Topological constraints in the LARGE-volume scenario}},
  \href{https://doi.org/10.1007/JHEP08(2022)226}{\emph{JHEP} {\bfseries 08}
  (2022) 226} [\href{https://arxiv.org/abs/2205.02856}{{\ttfamily
  2205.02856}}].

\bibitem{Blumenhagen:2022dbo}
R.~Blumenhagen, A.~Gligovic and S.~Kaddachi, \emph{{Mass Hierarchies and
  Quantum Gravity Constraints in DKMM-refined KKLT}},
  \href{https://arxiv.org/abs/2206.08400}{{\ttfamily 2206.08400}}.

\bibitem{Demirtas:2021nlu}
M.~Demirtas, M.~Kim, L.~McAllister, J.~Moritz and A.~Rios-Tascon, \emph{{Small
  cosmological constants in string theory}},
  \href{https://doi.org/10.1007/JHEP12(2021)136}{\emph{JHEP} {\bfseries 12}
  (2021) 136} [\href{https://arxiv.org/abs/2107.09064}{{\ttfamily
  2107.09064}}].

\bibitem{Demirtas:2021ote}
M.~Demirtas, M.~Kim, L.~McAllister, J.~Moritz and A.~Rios-Tascon,
  \emph{{Exponentially Small Cosmological Constant in String Theory}},
  \href{https://doi.org/10.1103/PhysRevLett.128.011602}{\emph{Phys. Rev. Lett.}
  {\bfseries 128} (2022) 011602}
  [\href{https://arxiv.org/abs/2107.09065}{{\ttfamily 2107.09065}}].

\bibitem{DeLuca:2022wfq}
G.~B. De~Luca, N.~De~Ponti, A.~Mondino and A.~Tomasiello, \emph{{Gravity from
  thermodynamics: optimal transport and negative effective dimensions}},
  \href{https://arxiv.org/abs/2212.02511}{{\ttfamily 2212.02511}}.

\bibitem{Acharya:2002kv}
B.~S. Acharya, \emph{{A Moduli fixing mechanism in M theory}},
  \href{https://arxiv.org/abs/hep-th/0212294}{{\ttfamily hep-th/0212294}}.

\bibitem{Petrini:2013ika}
M.~Petrini, G.~Solard and T.~Van~Riet, \emph{{AdS vacua with scale separation
  from IIB supergravity}},
  \href{https://doi.org/10.1007/JHEP11(2013)010}{\emph{JHEP} {\bfseries 11}
  (2013) 010} [\href{https://arxiv.org/abs/1308.1265}{{\ttfamily 1308.1265}}].

\bibitem{Caviezel:2009tu}
C.~Caviezel, T.~Wrase and M.~Zagermann, \emph{{Moduli Stabilization and
  Cosmology of Type IIB on SU(2)-Structure Orientifolds}},
  \href{https://doi.org/10.1007/JHEP04(2010)011}{\emph{JHEP} {\bfseries 04}
  (2010) 011} [\href{https://arxiv.org/abs/0912.3287}{{\ttfamily 0912.3287}}].

\bibitem{Caviezel:2008ik}
C.~Caviezel, P.~Koerber, S.~Kors, D.~Lust, D.~Tsimpis and M.~Zagermann,
  \emph{{The Effective theory of type IIA AdS(4) compactifications on
  nilmanifolds and cosets}},
  \href{https://doi.org/10.1088/0264-9381/26/2/025014}{\emph{Class. Quant.
  Grav.} {\bfseries 26} (2009) 025014}
  [\href{https://arxiv.org/abs/0806.3458}{{\ttfamily 0806.3458}}].

\bibitem{Richard:2014qsa}
J.-M. Richard, R.~Terrisse and D.~Tsimpis, \emph{{On the spin-2 Kaluza-Klein
  spectrum of $ {\mathrm{AdS}}_4\times
  {S}^2\left({\mathrm{\mathcal{B}}}_4\right) $}},
  \href{https://doi.org/10.1007/JHEP12(2014)144}{\emph{JHEP} {\bfseries 12}
  (2014) 144} [\href{https://arxiv.org/abs/1410.4669}{{\ttfamily 1410.4669}}].

\bibitem{Tsimpis:2012tu}
D.~Tsimpis, \emph{{Supersymmetric AdS vacua and separation of scales}},
  \href{https://doi.org/10.1007/JHEP08(2012)142}{\emph{JHEP} {\bfseries 08}
  (2012) 142} [\href{https://arxiv.org/abs/1206.5900}{{\ttfamily 1206.5900}}].

\bibitem{Buratti:2020kda}
G.~Buratti, J.~Calderon, A.~Mininno and A.~M. Uranga, \emph{{Discrete
  Symmetries, Weak Coupling Conjecture and Scale Separation in AdS Vacua}},
  \href{https://doi.org/10.1007/JHEP06(2020)083}{\emph{JHEP} {\bfseries 06}
  (2020) 083} [\href{https://arxiv.org/abs/2003.09740}{{\ttfamily
  2003.09740}}].

\bibitem{Lust:2020npd}
D.~L\"ust and D.~Tsimpis, \emph{{AdS$_{2}$ type-IIA solutions and scale
  separation}}, \href{https://doi.org/10.1007/JHEP07(2020)060}{\emph{JHEP}
  {\bfseries 07} (2020) 060}
  [\href{https://arxiv.org/abs/2004.07582}{{\ttfamily 2004.07582}}].

\bibitem{Font:2019uva}
A.~Font, A.~Herr\'aez and L.~E. Ib\'a\~nez, \emph{{On scale separation in type
  II AdS flux vacua}},
  \href{https://doi.org/10.1007/JHEP03(2020)013}{\emph{JHEP} {\bfseries 03}
  (2020) 013} [\href{https://arxiv.org/abs/1912.03317}{{\ttfamily
  1912.03317}}].

\bibitem{Basile:2022ypo}
I.~Basile, S.~Raucci and S.~Thom\'ee, \emph{{Revisiting Dudas-Mourad
  Compactifications}},
  \href{https://doi.org/10.3390/universe8100544}{\emph{Universe} {\bfseries 8}
  (2022) 544} [\href{https://arxiv.org/abs/2209.10553}{{\ttfamily
  2209.10553}}].

\bibitem{Marchesano:2020uqz}
F.~Marchesano, D.~Prieto, J.~Quirant and P.~Shukla, \emph{{Systematics of Type
  IIA moduli stabilisation}},
  \href{https://doi.org/10.1007/JHEP11(2020)113}{\emph{JHEP} {\bfseries 11}
  (2020) 113} [\href{https://arxiv.org/abs/2007.00672}{{\ttfamily
  2007.00672}}].

\bibitem{Emelin:2021gzx}
M.~Emelin, F.~Farakos and G.~Tringas, \emph{{Three-dimensional flux vacua from
  IIB on co-calibrated G2 orientifolds}},
  \href{https://doi.org/10.1140/epjc/s10052-021-09261-y}{\emph{Eur. Phys. J. C}
  {\bfseries 81} (2021) 456}
  [\href{https://arxiv.org/abs/2103.03282}{{\ttfamily 2103.03282}}].

\bibitem{Andriot:2022yyj}
D.~Andriot, L.~Horer and P.~Marconnet, \emph{{Exploring the landscape of
  (anti-) de Sitter and Minkowski solutions: group manifolds, stability and
  scale separation}},
  \href{https://doi.org/10.1007/JHEP08(2022)109}{\emph{JHEP} {\bfseries 08}
  (2022) 109} [\href{https://arxiv.org/abs/2204.05327}{{\ttfamily
  2204.05327}}].

\bibitem{Gautason:2015tig}
F.~F. Gautason, M.~Schillo, T.~Van~Riet and M.~Williams, \emph{{Remarks on
  scale separation in flux vacua}},
  \href{https://doi.org/10.1007/JHEP03(2016)061}{\emph{JHEP} {\bfseries 03}
  (2016) 061} [\href{https://arxiv.org/abs/1512.00457}{{\ttfamily
  1512.00457}}].

\bibitem{DeLuca:2021mcj}
G.~B. De~Luca and A.~Tomasiello, \emph{{Leaps and bounds towards scale
  separation}}, \href{https://doi.org/10.1007/JHEP12(2021)086}{\emph{JHEP}
  {\bfseries 12} (2021) 086}
  [\href{https://arxiv.org/abs/2104.12773}{{\ttfamily 2104.12773}}].

\bibitem{DeLuca:2021ojx}
G.~B. De~Luca, N.~De~Ponti, A.~Mondino and A.~Tomasiello, \emph{{Cheeger bounds
  on spin-two fields}},
  \href{https://doi.org/10.1007/JHEP12(2021)217}{\emph{JHEP} {\bfseries 12}
  (2021) 217} [\href{https://arxiv.org/abs/2109.11560}{{\ttfamily
  2109.11560}}].

\bibitem{McOrist:2012yc}
J.~McOrist and S.~Sethi, \emph{{M-theory and Type IIA Flux Compactifications}},
  \href{https://doi.org/10.1007/JHEP12(2012)122}{\emph{JHEP} {\bfseries 12}
  (2012) 122} [\href{https://arxiv.org/abs/1208.0261}{{\ttfamily 1208.0261}}].

\bibitem{Baines:2020dmu}
S.~Baines and T.~Van~Riet, \emph{{Smearing orientifolds in flux
  compactifications can be OK}},
  \href{https://doi.org/10.1088/1361-6382/aba8e0}{\emph{Class. Quant. Grav.}
  {\bfseries 37} (2020) 195015}
  [\href{https://arxiv.org/abs/2005.09501}{{\ttfamily 2005.09501}}].

\bibitem{Banks:2006hg}
T.~Banks and K.~van~den Broek, \emph{{Massive IIA flux compactifications and
  U-dualities}},
  \href{https://doi.org/10.1088/1126-6708/2007/03/068}{\emph{JHEP} {\bfseries
  03} (2007) 068} [\href{https://arxiv.org/abs/hep-th/0611185}{{\ttfamily
  hep-th/0611185}}].

\bibitem{Grimm:2004ua}
T.~W. Grimm and J.~Louis, \emph{{The Effective action of type IIA Calabi-Yau
  orientifolds}},
  \href{https://doi.org/10.1016/j.nuclphysb.2005.04.007}{\emph{Nucl. Phys. B}
  {\bfseries 718} (2005) 153}
  [\href{https://arxiv.org/abs/hep-th/0412277}{{\ttfamily hep-th/0412277}}].

\bibitem{Acharya:2006ne}
B.~S. Acharya, F.~Benini and R.~Valandro, \emph{{Fixing moduli in exact type
  IIA flux vacua}},
  \href{https://doi.org/10.1088/1126-6708/2007/02/018}{\emph{JHEP} {\bfseries
  02} (2007) 018} [\href{https://arxiv.org/abs/hep-th/0607223}{{\ttfamily
  hep-th/0607223}}].

\bibitem{Blaback:2010sj}
J.~Blaback, U.~H. Danielsson, D.~Junghans, T.~Van~Riet, T.~Wrase and
  M.~Zagermann, \emph{{Smeared versus localised sources in flux
  compactifications}},
  \href{https://doi.org/10.1007/JHEP12(2010)043}{\emph{JHEP} {\bfseries 12}
  (2010) 043} [\href{https://arxiv.org/abs/1009.1877}{{\ttfamily 1009.1877}}].

\bibitem{Marchesano:2020qvg}
F.~Marchesano, E.~Palti, J.~Quirant and A.~Tomasiello, \emph{{On supersymmetric
  AdS$_{4}$ orientifold vacua}},
  \href{https://doi.org/10.1007/JHEP08(2020)087}{\emph{JHEP} {\bfseries 08}
  (2020) 087} [\href{https://arxiv.org/abs/2003.13578}{{\ttfamily
  2003.13578}}].

\bibitem{Junghans:2020acz}
D.~Junghans, \emph{{O-Plane Backreaction and Scale Separation in Type IIA Flux
  Vacua}}, \href{https://doi.org/10.1002/prop.202000040}{\emph{Fortsch. Phys.}
  {\bfseries 68} (2020) 2000040}
  [\href{https://arxiv.org/abs/2003.06274}{{\ttfamily 2003.06274}}].

\bibitem{Marchesano:2022rpr}
F.~Marchesano, J.~Quirant and M.~Zatti, \emph{{New instabilities for
  non-supersymmetric AdS$_{4}$ orientifold vacua}},
  \href{https://doi.org/10.1007/JHEP10(2022)026}{\emph{JHEP} {\bfseries 10}
  (2022) 026} [\href{https://arxiv.org/abs/2207.14285}{{\ttfamily
  2207.14285}}].

\bibitem{Emelin:2022cac}
M.~Emelin, F.~Farakos and G.~Tringas, \emph{{O6-plane backreaction on
  scale-separated Type IIA AdS$_{3}$ vacua}},
  \href{https://doi.org/10.1007/JHEP07(2022)133}{\emph{JHEP} {\bfseries 07}
  (2022) 133} [\href{https://arxiv.org/abs/2202.13431}{{\ttfamily
  2202.13431}}].

\bibitem{Cribiori:2021gbf}
N.~Cribiori, D.~Lust and M.~Scalisi, \emph{{The gravitino and the swampland}},
  \href{https://doi.org/10.1007/JHEP06(2021)071}{\emph{JHEP} {\bfseries 06}
  (2021) 071} [\href{https://arxiv.org/abs/2104.08288}{{\ttfamily
  2104.08288}}].

\bibitem{Castellano:2021yye}
A.~Castellano, A.~Font, A.~Herraez and L.~E. Ib\'a\~nez, \emph{{A gravitino
  distance conjecture}},
  \href{https://doi.org/10.1007/JHEP08(2021)092}{\emph{JHEP} {\bfseries 08}
  (2021) 092} [\href{https://arxiv.org/abs/2104.10181}{{\ttfamily
  2104.10181}}].

\bibitem{DallAgata:2021nnr}
G.~Dall'Agata, M.~Emelin, F.~Farakos and M.~Morittu, \emph{{The unbearable
  lightness of charged gravitini}},
  \href{https://doi.org/10.1007/JHEP10(2021)076}{\emph{JHEP} {\bfseries 10}
  (2021) 076} [\href{https://arxiv.org/abs/2108.04254}{{\ttfamily
  2108.04254}}].

\bibitem{Ooguri:2006in}
H.~Ooguri and C.~Vafa, \emph{{On the Geometry of the String Landscape and the
  Swampland}},
  \href{https://doi.org/10.1016/j.nuclphysb.2006.10.033}{\emph{Nucl. Phys. B}
  {\bfseries 766} (2007) 21}
  [\href{https://arxiv.org/abs/hep-th/0605264}{{\ttfamily hep-th/0605264}}].

\bibitem{Etheredge:2022opl}
M.~Etheredge, B.~Heidenreich, S.~Kaya, Y.~Qiu and T.~Rudelius,
  \emph{{Sharpening the Distance Conjecture in Diverse Dimensions}},
  \href{https://arxiv.org/abs/2206.04063}{{\ttfamily 2206.04063}}.

\bibitem{Kachru:2002gs}
S.~Kachru, J.~Pearson and H.~L. Verlinde, \emph{{Brane / flux annihilation and
  the string dual of a nonsupersymmetric field theory}},
  \href{https://doi.org/10.1088/1126-6708/2002/06/021}{\emph{JHEP} {\bfseries
  06} (2002) 021} [\href{https://arxiv.org/abs/hep-th/0112197}{{\ttfamily
  hep-th/0112197}}].

\bibitem{Scalisi:2020jal}
M.~Scalisi, P.~Soler, V.~Van~Hemelryck and T.~Van~Riet, \emph{{Conifold
  dynamics and axion monodromies}},
  \href{https://doi.org/10.1007/JHEP10(2020)133}{\emph{JHEP} {\bfseries 10}
  (2020) 133} [\href{https://arxiv.org/abs/2007.15391}{{\ttfamily
  2007.15391}}].

\bibitem{Lust:2004ig}
D.~Lust and D.~Tsimpis, \emph{{Supersymmetric AdS(4) compactifications of IIA
  supergravity}},
  \href{https://doi.org/10.1088/1126-6708/2005/02/027}{\emph{JHEP} {\bfseries
  02} (2005) 027} [\href{https://arxiv.org/abs/hep-th/0412250}{{\ttfamily
  hep-th/0412250}}].

\bibitem{Grana:2006kf}
M.~Grana, R.~Minasian, M.~Petrini and A.~Tomasiello, \emph{{A Scan for new N=1
  vacua on twisted tori}},
  \href{https://doi.org/10.1088/1126-6708/2007/05/031}{\emph{JHEP} {\bfseries
  05} (2007) 031} [\href{https://arxiv.org/abs/hep-th/0609124}{{\ttfamily
  hep-th/0609124}}].

\bibitem{Behrndt:2004mj}
K.~Behrndt and M.~Cvetic, \emph{{General N=1 supersymmetric fluxes in massive
  type IIA string theory}},
  \href{https://doi.org/10.1016/j.nuclphysb.2004.12.004}{\emph{Nucl. Phys. B}
  {\bfseries 708} (2005) 45}
  [\href{https://arxiv.org/abs/hep-th/0407263}{{\ttfamily hep-th/0407263}}].

\bibitem{Grana:2004bg}
M.~Grana, R.~Minasian, M.~Petrini and A.~Tomasiello, \emph{{Supersymmetric
  backgrounds from generalized Calabi-Yau manifolds}},
  \href{https://doi.org/10.1088/1126-6708/2004/08/046}{\emph{JHEP} {\bfseries
  08} (2004) 046} [\href{https://arxiv.org/abs/hep-th/0406137}{{\ttfamily
  hep-th/0406137}}].

\bibitem{Gautason:2015tla}
F.~F. Gautason, B.~Truijen and T.~Van~Riet, \emph{{The many faces of brane-flux
  annihilation}}, \href{https://doi.org/10.1007/JHEP10(2015)152}{\emph{JHEP}
  {\bfseries 10} (2015) 152}
  [\href{https://arxiv.org/abs/1505.00159}{{\ttfamily 1505.00159}}].

\bibitem{Narayan:2010em}
P.~Narayan and S.~P. Trivedi, \emph{{On The Stability Of Non-Supersymmetric AdS
  Vacua}}, \href{https://doi.org/10.1007/JHEP07(2010)089}{\emph{JHEP}
  {\bfseries 07} (2010) 089} [\href{https://arxiv.org/abs/1002.4498}{{\ttfamily
  1002.4498}}].

\bibitem{Marchesano:2021ycx}
F.~Marchesano, D.~Prieto and J.~Quirant, \emph{{BIonic membranes and AdS
  instabilities}}, \href{https://doi.org/10.1007/JHEP07(2022)118}{\emph{JHEP}
  {\bfseries 07} (2022) 118}
  [\href{https://arxiv.org/abs/2110.11370}{{\ttfamily 2110.11370}}].

\bibitem{Klebanov:2000hb}
I.~R. Klebanov and M.~J. Strassler, \emph{{Supergravity and a confining gauge
  theory: Duality cascades and chi SB resolution of naked singularities}},
  \href{https://doi.org/10.1088/1126-6708/2000/08/052}{\emph{JHEP} {\bfseries
  08} (2000) 052} [\href{https://arxiv.org/abs/hep-th/0007191}{{\ttfamily
  hep-th/0007191}}].

\bibitem{reid1987moduli}
M.~Reid, \emph{The moduli space of 3-folds withk= 0 may nevertheless be
  irreducible}, {\emph{Mathematische Annalen} {\bfseries 278} (1987) 329}.

\bibitem{Silverstein:2008sg}
E.~Silverstein and A.~Westphal, \emph{{Monodromy in the CMB: Gravity Waves and
  String Inflation}},
  \href{https://doi.org/10.1103/PhysRevD.78.106003}{\emph{Phys. Rev. D}
  {\bfseries 78} (2008) 106003}
  [\href{https://arxiv.org/abs/0803.3085}{{\ttfamily 0803.3085}}].

\bibitem{McAllister:2008hb}
L.~McAllister, E.~Silverstein and A.~Westphal, \emph{{Gravity Waves and Linear
  Inflation from Axion Monodromy}},
  \href{https://doi.org/10.1103/PhysRevD.82.046003}{\emph{Phys. Rev. D}
  {\bfseries 82} (2010) 046003}
  [\href{https://arxiv.org/abs/0808.0706}{{\ttfamily 0808.0706}}].

\bibitem{Marchesano:2014mla}
F.~Marchesano, G.~Shiu and A.~M. Uranga, \emph{{F-term Axion Monodromy
  Inflation}}, \href{https://doi.org/10.1007/JHEP09(2014)184}{\emph{JHEP}
  {\bfseries 09} (2014) 184} [\href{https://arxiv.org/abs/1404.3040}{{\ttfamily
  1404.3040}}].

\bibitem{Blumenhagen:2014gta}
R.~Blumenhagen and E.~Plauschinn, \emph{{Towards Universal Axion Inflation and
  Reheating in String Theory}},
  \href{https://doi.org/10.1016/j.physletb.2014.08.007}{\emph{Phys. Lett. B}
  {\bfseries 736} (2014) 482}
  [\href{https://arxiv.org/abs/1404.3542}{{\ttfamily 1404.3542}}].

\bibitem{Hebecker:2014eua}
A.~Hebecker, S.~C. Kraus and L.~T. Witkowski, \emph{{D7-Brane Chaotic
  Inflation}},
  \href{https://doi.org/10.1016/j.physletb.2014.08.028}{\emph{Phys. Lett. B}
  {\bfseries 737} (2014) 16} [\href{https://arxiv.org/abs/1404.3711}{{\ttfamily
  1404.3711}}].

\bibitem{Brown:2016nqt}
J.~Brown, W.~Cottrell, G.~Shiu and P.~Soler, \emph{{Tunneling in Axion
  Monodromy}}, \href{https://doi.org/10.1007/JHEP10(2016)025}{\emph{JHEP}
  {\bfseries 10} (2016) 025}
  [\href{https://arxiv.org/abs/1607.00037}{{\ttfamily 1607.00037}}].

\bibitem{Hebecker:2022zme}
A.~Hebecker, S.~Schreyer and G.~Venken, \emph{{Curvature corrections to KPV: do
  we need deep throats?}},
  \href{https://doi.org/10.1007/JHEP10(2022)166}{\emph{JHEP} {\bfseries 10}
  (2022) 166} [\href{https://arxiv.org/abs/2208.02826}{{\ttfamily
  2208.02826}}].

\bibitem{Bachas:1999um}
C.~P. Bachas, P.~Bain and M.~B. Green, \emph{{Curvature terms in D-brane
  actions and their M theory origin}},
  \href{https://doi.org/10.1088/1126-6708/1999/05/011}{\emph{JHEP} {\bfseries
  05} (1999) 011} [\href{https://arxiv.org/abs/hep-th/9903210}{{\ttfamily
  hep-th/9903210}}].

\bibitem{Junghans:2014zla}
D.~Junghans and G.~Shiu, \emph{{Brane curvature corrections to the $
  \mathcal{N} =$ 1 type II/F-theory effective action}},
  \href{https://doi.org/10.1007/JHEP03(2015)107}{\emph{JHEP} {\bfseries 03}
  (2015) 107} [\href{https://arxiv.org/abs/1407.0019}{{\ttfamily 1407.0019}}].

\bibitem{Burgess:2006mn}
C.~P. Burgess, P.~G. Camara, S.~P. de~Alwis, S.~B. Giddings, A.~Maharana,
  F.~Quevedo et~al., \emph{{Warped Supersymmetry Breaking}},
  \href{https://doi.org/10.1088/1126-6708/2008/04/053}{\emph{JHEP} {\bfseries
  04} (2008) 053} [\href{https://arxiv.org/abs/hep-th/0610255}{{\ttfamily
  hep-th/0610255}}].

\bibitem{Gratton:2000fj}
S.~Gratton and N.~Turok, \emph{{Homogeneous modes of cosmological instantons}},
  \href{https://doi.org/10.1103/PhysRevD.63.123514}{\emph{Phys. Rev. D}
  {\bfseries 63} (2001) 123514}
  [\href{https://arxiv.org/abs/hep-th/0008235}{{\ttfamily hep-th/0008235}}].

\bibitem{Myers:1999ps}
R.~C. Myers, \emph{{Dielectric branes}},
  \href{https://doi.org/10.1088/1126-6708/1999/12/022}{\emph{JHEP} {\bfseries
  12} (1999) 022} [\href{https://arxiv.org/abs/hep-th/9910053}{{\ttfamily
  hep-th/9910053}}].

\bibitem{Cascales:2004qp}
J.~F.~G. Cascales and A.~M. Uranga, \emph{{Branes on generalized calibrated
  submanifolds}},
  \href{https://doi.org/10.1088/1126-6708/2004/11/083}{\emph{JHEP} {\bfseries
  11} (2004) 083} [\href{https://arxiv.org/abs/hep-th/0407132}{{\ttfamily
  hep-th/0407132}}].

\bibitem{Marchesano:2006ns}
F.~Marchesano, \emph{{D6-branes and torsion}},
  \href{https://doi.org/10.1088/1126-6708/2006/05/019}{\emph{JHEP} {\bfseries
  05} (2006) 019} [\href{https://arxiv.org/abs/hep-th/0603210}{{\ttfamily
  hep-th/0603210}}].

\end{thebibliography}\endgroup

\end{document}